\documentclass[nofootinbib,superscriptaddress,a4paper,twocolumn]{revtex4-1}
\usepackage{geometry}
\usepackage[english]{babel}
\usepackage[latin1]{inputenc}
\usepackage{hyperref}
\usepackage{amsmath, amssymb, amsfonts, mathrsfs}
\usepackage{graphicx}
\usepackage{color}

\newcommand{\ket}[1]{|#1\rangle}

\begin{document}

\title{Quantum Experiments and Graphs II:\\Quantum Interference, Computation and State Generation}

\author{Xuemei Gu}
\email{xmgu@smail.nju.edu.cn}
\affiliation{Institute for Quantum Optics and Quantum Information (IQOQI), Austrian Academy of Sciences, Boltzmanngasse 3, 1090 Vienna, Austria.}
\affiliation{State Key Laboratory for Novel Software Technology, Nanjing University, 163 Xianlin Avenue, Qixia District, 210023, Nanjing City, China.}

\author{Manuel Erhard}
\affiliation{Institute for Quantum Optics and Quantum Information (IQOQI), Austrian Academy of Sciences, Boltzmanngasse 3, 1090 Vienna, Austria.}
\affiliation{Vienna Center for Quantum Science \& Technology (VCQ), Faculty of Physics, University of Vienna, Boltzmanngasse 5, 1090 Vienna, Austria.}

\author{Anton Zeilinger}
\email{anton.zeilinger@univie.ac.at}
\affiliation{Institute for Quantum Optics and Quantum Information (IQOQI), Austrian Academy of Sciences, Boltzmanngasse 3, 1090 Vienna, Austria.}
\affiliation{Vienna Center for Quantum Science \& Technology (VCQ), Faculty of Physics, University of Vienna, Boltzmanngasse 5, 1090 Vienna, Austria.}

\author{Mario Krenn}
\email{mario.krenn@univie.ac.at; present address: Department of Chemistry, University of Toronto, Toronto, Ontario M5S 3H6, Canada.}
\affiliation{Institute for Quantum Optics and Quantum Information (IQOQI), Austrian Academy of Sciences, Boltzmanngasse 3, 1090 Vienna, Austria.}
\affiliation{Vienna Center for Quantum Science \& Technology (VCQ), Faculty of Physics, University of Vienna, Boltzmanngasse 5, 1090 Vienna, Austria.}

\begin{abstract}
We present a conceptually new approach to describe state-of-the-art photonic quantum experiments using Graph Theory. There, the quantum states are given by the coherent superpositions of perfect matchings. The crucial observation is that introducing complex weights in graphs naturally leads to quantum interference. The new viewpoint immediately leads to many interesting results, some of which we present here. Firstly, we identify a new and experimentally completely unexplored multiphoton interference phenomenon. Secondly, we find that computing the results of such experiments is \textsc{\#P}-hard, which means it is a classically intractable problem dealing with the computation of a matrix function \textit{Permanent} and its generalization \textit{Hafnian}. Thirdly, we explain how a recent no-go result applies generally to linear optical quantum experiments, thus revealing important insights to quantum state generation with current photonic technology. Fourthly, we show how to describe quantum protocols such as entanglement swapping in a graphical way. The uncovered bridge between quantum experiments and Graph Theory offers a novel perspective on a widely used technology, and immediately raises many follow-up questions.
\end{abstract}

\date{\today}
\maketitle

Photonic quantum experiments prominently use probabilistic photon sources in combination with linear optics \cite{pan2012multiphoton}. This allows for the generation of multipartite quantum entanglement such as Greenberger-Horne-Zeilinger (GHZ) states \cite{bouwmeester1999observation, yao2012observation, wang2016experimental, wang201818}, W states \cite{eibl2004experimental}, Dicke states \cite{wieczorek2009experimental, hiesmayr2016observation} or high-dimensional states \cite{malik2016multi, erhard2018experimental}, proof-of-principle experiments of special-purpose quantum computing \cite{aaronson2011computational, rahimi2015can, lund2017quantum, spring2012boson, broome2013photonic, tillmann2013experimental, crespi2013integrated, carolan2015universal} or applications such as quantum teleportation \cite{bouwmeester1997experimental, wang2015quantum} and entanglement swapping \cite{pan1998experimental, zhang2017simultaneous}.

Here we show that one can describe all of these quantum experiments\footnote{The experiments mentioned before all consist of probabilistic photon pair sources and linear optics. This is what we mean by \textit{quantum experiments} for the rest of the manuscript. We show that graphs can describe all of such experiments. Additionally, the property of perfect matchings corresponds to N-fold coincidence detection, which is widely used in quantum optics experiments.} with graph theory. To do this, we generalize a recently found link between graphs and a special type of quantum experiments with multiple crystals \cite{krenn2017quantum} -- which were based on the computer-inspired concept of \textit{Entanglement by Path Identity} \cite{krenn2016automated, krenn2017entanglement}. By introducing complex weights in graphs, we can naturally describe the operations of linear optical elements, such as phase shifters and beam splitters, which enables us to describe quantum interference effects. This technique allows us to find several results:
\begin{figure}[!t]
\centering
\includegraphics[width=\linewidth]{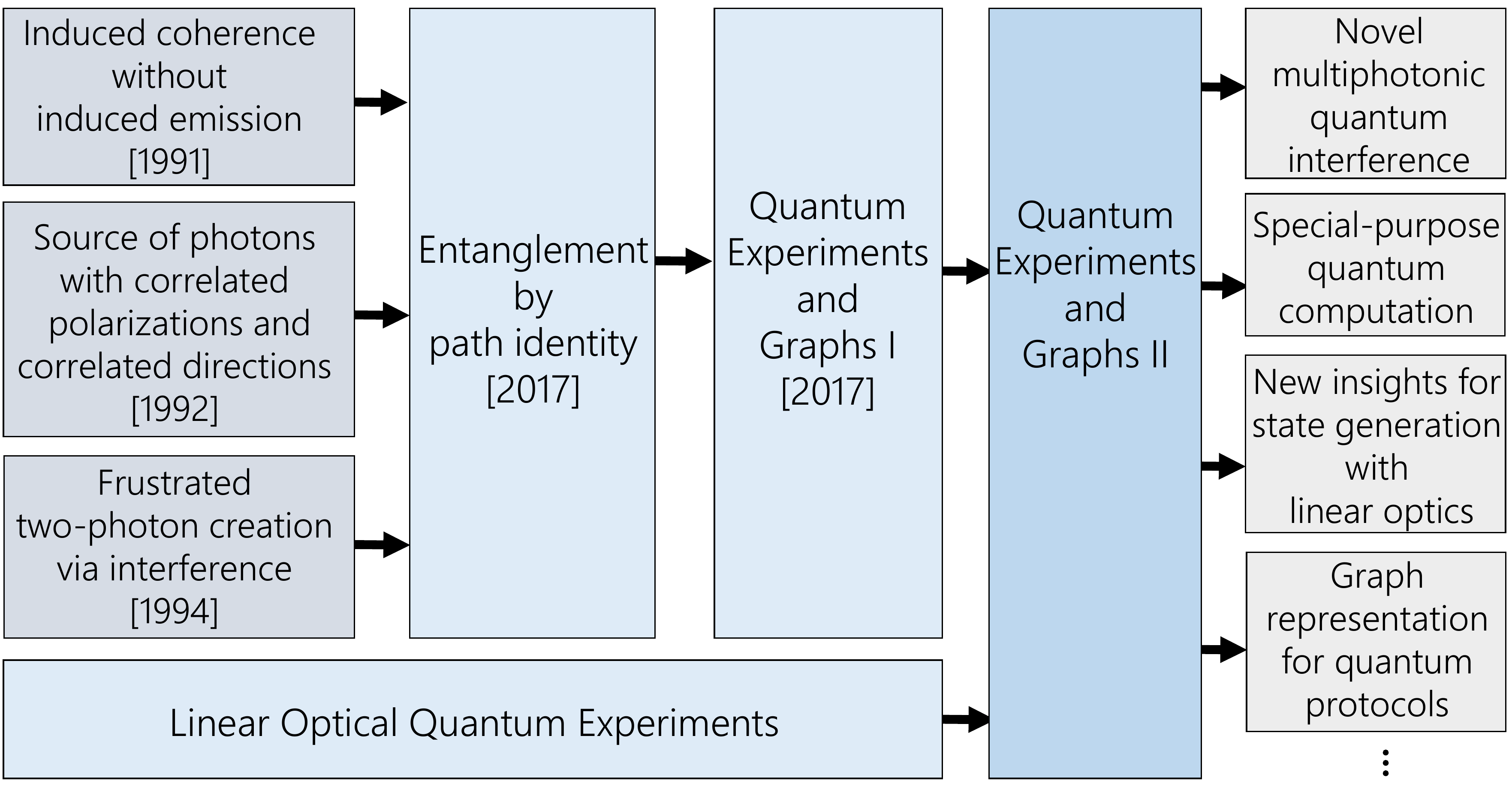}
\caption{A rough sketch of the influences that have led to the current manuscript. Three seminal papers \cite{wang1991induced,herzog1994frustrated,hardy1992source} have influence Entanglement by Path Identity \cite{krenn2017entanglement}, which itself has led to Quantum Experiments and Graphs I \cite{krenn2017quantum}. Here we connect these ideas with the mature field of research that investigate passive linear optics in the quantum regime. The results of the merger are described in the current manuscript.}
\label{fig:fig1MainPoint}
\end{figure}
(1) We identify a novel multiphotonic quantum interference effect which is based on generalization of frustrated pair-creation\footnote{Frustrated pair-creation is an effect where the amplitudes of two pair-creation events can constructively or destructively interfere.} in a network of nonlinear crystals. Although the two-photon special case of this interference effect has been observed more than 20 years ago \cite{herzog1994frustrated}, the multiphoton generalisation with many crystals has neither been investigated theoretically nor experimentally before. (2) We find these networks of crystals cannot be calculated efficiently on a classical computer. The experimental output distributions require the summation of weights of perfect matchings\footnote{The weight of a perfect matching is the product of the weights of all containing edges.} in a complex weighted graph (or alternatively, probabilities proportional to the matrix function \textit{Permanent} and its generalization \textit{Hafnian}), which is \textsc{\#P}-hard \cite{valiant1979complexity, aaronson2011linear}\footnote{A \textsc{\#P}-hard problem is at least as difficult as any problem in the complexity class \textsc{\#P}.} -- and related to the \textit{BosonSampling} problem. (3) We show that insights from graph theory identify restrictions on the possibility of realizing certain classes of entangled states with current photonic technology. (4) The graph-theoretical description of experiments also leads to a pictorial explanation of quantum protocols such as entanglement swapping. We expect that this will help in designing or intuitively understanding novel (high-dimensional) quantum protocols. The conceptual ideas that have led to this article are shown in Fig. \ref{fig:fig1MainPoint}.

Connections between graph theory and quantum physics have been drawn in earlier complementary works. A well-known example is the so-called graph states, which can be used for universal quantum computation \cite{raussendorf2001one, hein2006entanglement}. That approach has been generalized to continuous-variable quantum computation \cite{menicucci2006universal}, using an interesting connection between gaussian states and graphs \cite{menicucci2011graphical}. Graphs have also been used to study collective phases of quantum systems \cite{shchesnovich2018collective} and used to investigate Quantum Random Networks \cite{perseguers2009quantum, cuquet2009entanglement}. The bridge between graphs and quantum experiments that we present here is quite different, thus allowing us to explore entirely different questions. The correspondence between graph theory and quantum experiments is listed in Table. \ref{tab:compare}.

\begin{table}[t]
  \centering
  \caption{The analogies for Quantum Experiments and Graph Theory.}
    \begin{tabular}{p{4.5cm}|p{2.8cm}}
    \hline
    \textbf{Linear Optical Quantum Experiments}&\textbf{Graph Theory}\\ \hline
    Quantum photonic setup including linear optical elements and non-linear crystals as probabilistic photon pair sources& Complex weighted undirected Graph \\ \hline
    optical output path& \textit{vertex set S}\\ \hline
    photonic modes in optical output path& vertices in \textit{vertex set S} \\ \hline
    mode numbers& labels of the vertices\\ \hline
    photon pair correlation &Edges\\ \hline
    phase between photonic modes &color of the edges \\ \hline
    amplitude of photonic modes &width of the edges \\ \hline
    n-fold coincidence& perfect matching \\ \hline
    \#(terms in quantum state)&\#(perfect matchings)\\ \hline
    \end{tabular}
    \label{tab:compare}
\end{table}

\begin{figure}[!t]
\centering
\includegraphics[width=0.98\linewidth]{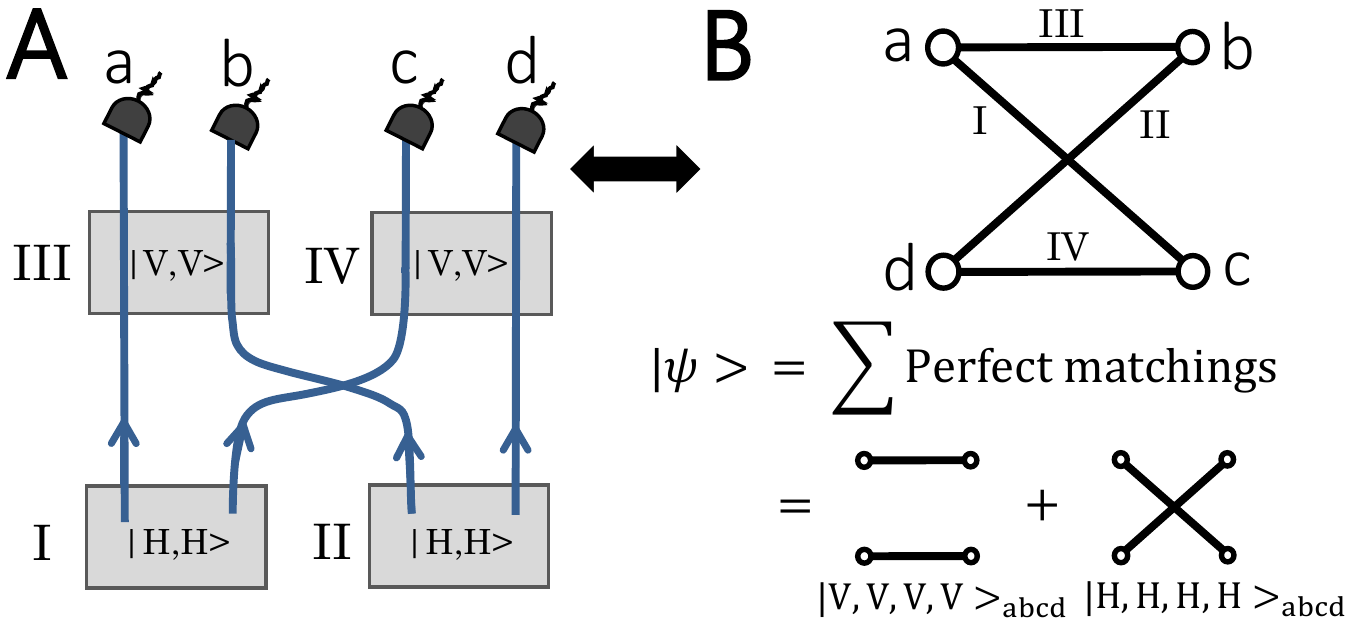}
\caption{Generation of 2-dimensional 4-photon GHZ state using \textit{Entanglement by Path Identity} \cite{krenn2017entanglement} and corresponding graph description of the setup \cite{krenn2017quantum}. \textbf{A}: An optical setup consists of four probabilistic photon pair sources, for example non-linear crystals. The crystals (gray squares) I-IV are pumped coherently and the pump power is set in such a way that two photon pairs are produced. Here we take the polarization for simplicity -- crystals I and II each produces photon pair with $\ket{H,H}$ while crystals III and IV create photon pair with $\ket{V,V}$. The four-fold coincidence requires a photon in each detector simultaneously, which can only happen when crystals I and II or crystals III and IV fire together. \textbf{B}: The corresponding graph of the experiment. Each vertex stands for a photon path and each edge represents one crystal. Thus the graph has four vertices and four edges. The condition of four-fold coincidence is represented by the \textit{perfect matchings of the graph -- a subset of edges that contains every vertex exactly once}. There are two subsets of edges ($E_{ab}, E_{dc}$) and ($E_{ac}, E_{bd}$) which form the perfect matchings in the graph. The final output state with post-selection is in a superposition of all the possibilities. Therefore, it can be seen as a superposition of all the perfect matchings of the graph, which gives the result $\ket{\psi}=\frac{1}{\sqrt{2}}(\ket{H,H,H,H}_{abcd}+\ket{V,V,V,V}_{abcd})$.}
\label{fig:fig2PathGraph}
\end{figure}
\section*{Entanglement by Path Identity and Graphs}

In this section, we briefly explain the main ideas from \textit{Entanglement by Path Identity} \cite{krenn2017entanglement} and \textit{Quantum Experiments and Graphs I} \cite{krenn2017quantum}, which form the basis for the rest of this manuscript. The concept of \textit{Entanglement by Path Identity} shows a new and very general way to experimentally produce multipartite and high-dimensional entanglement. Such type of experiments can be translated into graphs \cite{krenn2017quantum}. As an example, we show an experimental setup which creates a two-dimensional GHZ state in polarization, see Fig. \ref{fig:fig2PathGraph}A. The probabilistic photon pair sources (for example, the nonlinear crystals) are set up in such a way that crystals I and II can create horizontally polarized photon pairs, while crystals III and IV produce vertically polarized photon pairs. All the crystals are excitated coherently and the laser pump power is set such that two photon pairs are produced simultaneously.\footnote{The pair creation process of SPDC is entirely probabilistic. That means, the probability that two pairs are created in one single crystals is as high as the creation of two pairs in two crystals. That furthermore means that if creating one pair of photons has a probability of $p$, then creating two pairs has the probability $p^2$. In the experiment depicted in Fig. \ref{fig:fig2PathGraph}A, with some probability, more than two pairs are created. These higher-order photon pairs are the main source of reduced fidelity in multi-photonic GHZ state experiments \cite{wang2016experimental}. However, the laser power can be adjusted such that these cases have a sufficiently low probability (of course, with the drawback of lower count rates). The same arguments hold for all other examples in the manuscript (as they do for most other SPDC-based quantum optics experiments).}

The final state is obtained under the condition of four-fold coincidences, which means that all four detectors click simultaneously. \footnote{Most multi-photonic entangled quantum states are created under the condition of N-fold coincidence detection \cite{pan2012multiphoton}. It allows for investigation and application of these states as long as the photon paths are not combined anymore, such as in a subsequent linear optical setup. In that case, one needs to analyse the perfect matchings after the entire setup. Alternatively, one can use a photon number filter based on quantum teleportation in each output of the setup, as introduced in \cite{wang2015quantum}.} This can only happen if the two photon pairs origin either from crystals I and II or from crystals III and IV. There is no other case to fulfill the four-fold coincidence condition. For example, if crystal I and III fire together, there is no photon in path $d$, while there are two photons in path $a$. The final quantum state after post-selection can thus be written as $\ket{\psi} = \frac{1}{\sqrt{2}}(\ket{H,H,H,H}_{abcd}+\ket{V,V,V,V}_{abcd})$, where H and V stand for horizontal and vertical polarization respectively, and the subscripts $a$, $b$, $c$ and $d$ represent the photon's paths.

One can describe such types of quantum experiments using graph theory \cite{krenn2017quantum}. There, each vertex represents a photon path and each edge stands for a nonlinear crystal which can probabilistically produce a correlated photon pair. Therefore, the experiment can be described with a graph of four vertices and four edges depicted in Fig. \ref{fig:fig2PathGraph}B. A four-fold coincidence is given by a \textit{perfect matching of the graph, which is a subset of edges that contains every vertex exactly once}. For example, there are two subsets of edges ($E_{ab}$, $E_{dc}$) and ($E_{ac}$, $E_{bd}$) in Fig. \ref{fig:fig2PathGraph}B, which form the two perfect matchings. Thus, the final quantum state after post-selection can be seen as the coherent superposition of all perfect matchings of the graph.

\begin{figure}[!t]
\centering
\includegraphics[width=\linewidth]{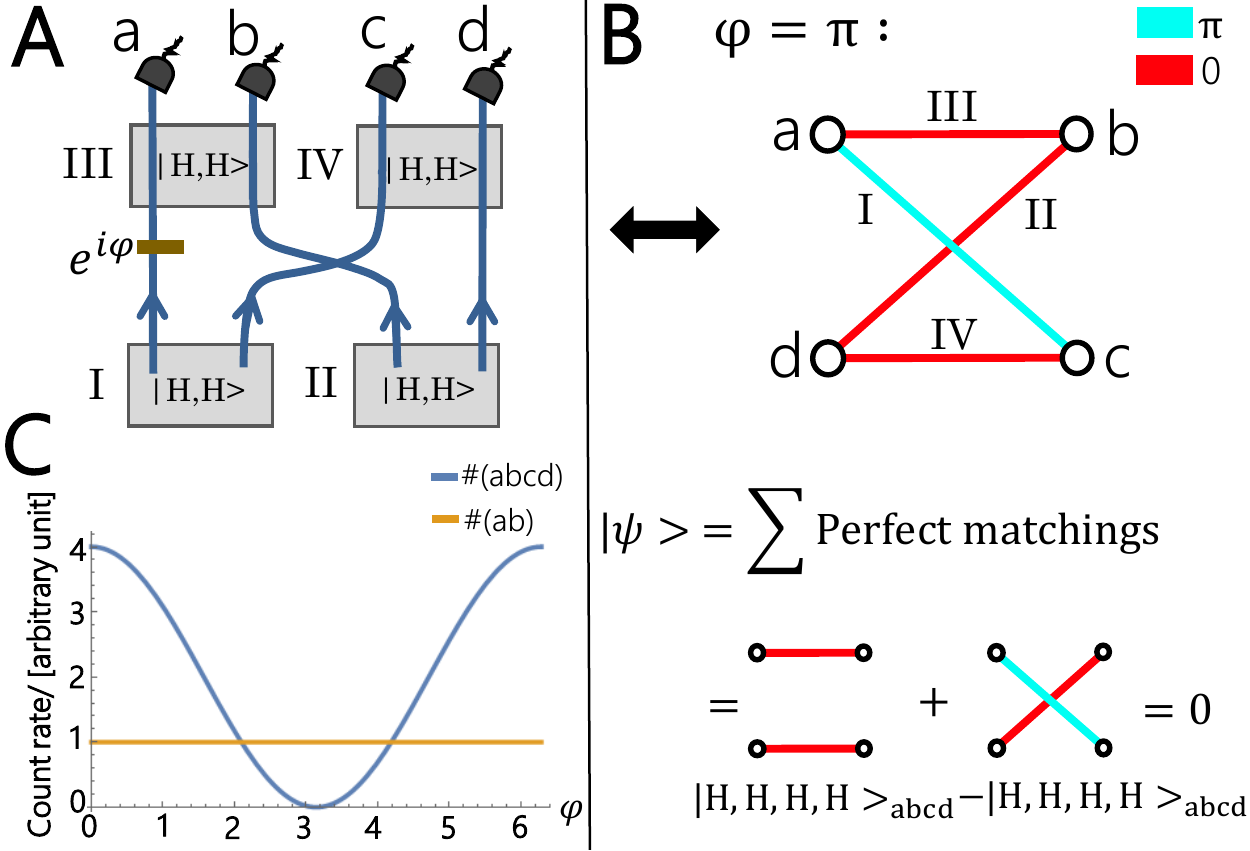}
\caption{Interference of \textit{perfect matchings}. \textbf{A}: A setup with all crystals producing horizontally polarized photon pairs. A phase shifter with a phase of $\varphi$ is inserted between crystals I and III. \textbf{B}: The corresponding graph of the experimental setup. The complex weight $e^{i\varphi}$ introduced by the phase shifter ($e^{i\varphi}$, here: $\varphi=\pi$) is depicted with different colors. Here red and blue of the edge stand for $0$ and $\pi$ phase shift. There are two perfect matchings of the graph, which come from crystals I and II and crystals III and IV, respectively. When one calculates the sum of the perfect matchings, the quantum state is given by $\ket{\psi}=(1+e^{i\pi})\ket{H,H,H,H}_{abcd}=0$. This means the two perfect matchings cancel each other. \textbf{C}: When the phase $\varphi$ changes from $0$ to $2\pi$, one can see the 4-fold coincidence (depicted as $\#(abcd)$) count rate changes while the 2-fold coincidence (for example, number of photon pairs in outputs $a$ and $b$, depicted as $\#(ab)$) count rate remains constant.}
\label{fig:fig3ComplexWeigth}
\end{figure}

\begin{figure*}[!t]
\centering
\includegraphics[width=\linewidth]{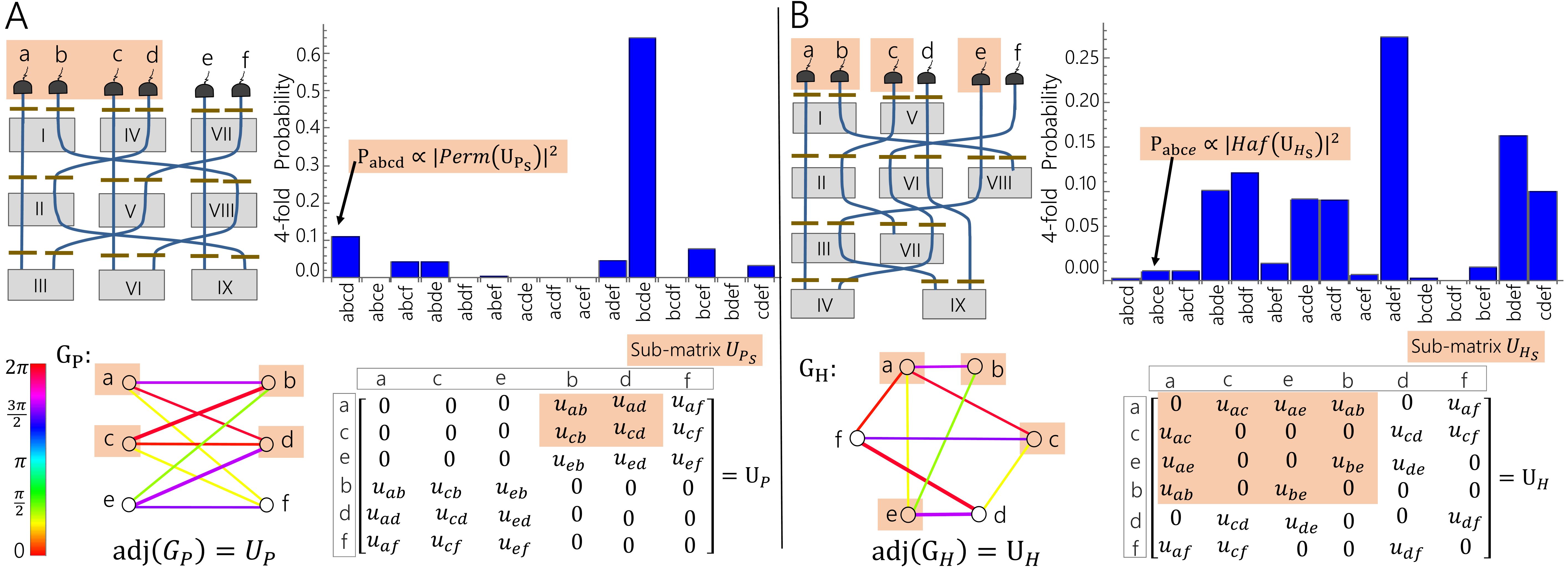}
\caption{Quantum experiments and Computation complexity. \textbf{A}: An experiment consisting of 9 nonlinear crystals (with labels I-IX) and 18 phase shifters (gold lines). They are arranged such that paths $a$, $c$ and $e$ are parallel. All the crystals are pumped coherently and can produce indistinguishable photon pairs. The pump power is set in such a way that two crystals can produce photon pairs. One can adjust the phase shifters and pump power to change the phases and transition amplitudes (the values are shown in SI Appendix \cite{supp}). The corresponding graph $G_{P}$ and its adjacency matrix $adj(G_{P})$ for the setup are at the bottom. The ordering of the column and row are ($a$, $c$, $e$, $b$, $d$ and $f$). Calculating four-fold coincidences in one specific subset path ($a$, $b$, $c$ and $d$) of four outputs relates to summing the weights of perfect matchings of the sub-graph with related vertices, which corresponds to computing the matrix function \textit{Permanent} of sub-matrix $U_{P_{s}}$ highlighted in orange. Thus, the probability that a certain arrangement of detectors click $P_{abcd}$ is proportional to the $|Perm(U_{P_{s}})|^2$. All the combinations for the four-fold coincidence are depicted in the histogram (details see SI Appendix \cite{supp}). \textbf{B}: A crystal network that shows the general case. The 9 crystals and 18 phase shifters are randomly put in order. In analogous to \textbf{A}, the pump power is also set such that two photon pairs can be created. The lower part shows the corresponding graph $G_{H}$ and its adjacency matrix $adj(G_{H})$, where the ordering is the same as $U_{P}$. Again, we calculate the four-fold coincidence in specific outputs $a$, $b$, $c$ and $e$. This corresponds to computing the \textit{Hafnian} of sub-matrix $U_{H_{s}}$, which is a generalisation of the \textit{Permanent}. The probability $P_{abce}$ is given by the matrix function \textit{Hafnian}, $P_{abce} \propto |Haf(U_{H_{s}})|^2$.}
\label{fig:fig4BosonSampling}
\end{figure*}
\section*{Complex weighted Graphs -- Quantum Experiments}

\subsection*{Quantum Interference}

Now we start generalising the connection between quantum experiments and graphs. The crucial observation is that one can deal with a phase shifter in the quantum experiment as a complex weight in the graph. When we add phase shifters in the experiments and all the crystals produce indistinguishable photon pairs, the experimental output probability with four-fold post-selection is given by the superposition of the perfect matchings of the graph weighted with a complex number.

As an example shown in Fig. \ref{fig:fig3ComplexWeigth}A, we insert a phase shifter between crystals I and III and all the four crystals create horizontally polarized photon pairs. The phase $\varphi$ is set to a phase shift of $\pi$ and the pump power is set such that two photon pairs are created. With the graph-experimental connection, one can also describe the experimental setup as a graph which is depicted in Fig. \ref{fig:fig3ComplexWeigth}B. The color of the edge stands for the phase in the experiments while the width of the edge represents the absolute value of the amplitude. In order to calculate four-fold coincidences from the outputs, we need to sum the weights of perfect matchings of the corresponding graph. There are two perfect matchings of the graph, where one is given by crystals III and IV while the other is from crystal I and II. The interference of the two perfect matchings (which means, of the two four-fold possibilities) can be obtained by varying the relative complex weight $e^{i\varphi}$ between them. Therefore, the cancellation of the perfect matchings shows the destructive interference in the experiment.

More quantitatively, each nonlinear crystal probabilistically creates photon pairs from spontaneous parametric down-conversion (SPDC). We follow the theoretical method presented in \cite{wang1991induced,lahiri2015theory}, and describe the down-conversion creation process as
\begin{equation}
\begin{aligned}
\hat{U}\approx 1+g(\hat{a}^{\dag}\hat{b}^{\dag})+\frac{g^2}{2}(\hat{a}^{\dag}\hat{b}^{\dag})^2+O(g^3)
\end{aligned}
\end{equation}

where $\hat{a}^{\dag}$ and $\hat{b}^{\dag}$ are single-photon creation operators in paths $a$ and $b$, and $g$ is the down-conversion amplitude. The terms of $O(g^3)$ and higher are neglected. The quantum state can be expressed as $\ket{\psi}=\hat{U}\ket{vac}$, where $\ket{vac}$ is the vacuum state.

Here we neglect the empty modes and higher-order terms, and only write first order terms and the four-fold term for second order spontaneous parametric down-conversion. The full state up to second order can be see in SI Appendix \cite{supp}. Therefore, the final quantum state in our example is
\begin{equation}
\begin{aligned}
\ket{\psi}=g(\ket{H,H}_{ab}+\ket{H,H}_{cd}+\ket{H,H}_{bd}+e^{i\varphi}\ket{H,H}_{ac})\\
+g^2(1+e^{i\varphi})\ket{H,H,H,H}_{abcd}+...
\end{aligned}
\end{equation}

We can see that the four-fold coincidence count rate varies with the tunable phase $\varphi$ while the two-fold coincidence count rate remains constant, which is depicted in Fig. \ref{fig:fig3ComplexWeigth}C. This is a multiphotonic generalization of two photon frustrated down-conversion \cite{herzog1994frustrated} that has never been experimentally observed.

\subsection*{Special-purpose quantum computation}

We here show a generalization of the setup in Fig. \ref{fig:fig3ComplexWeigth}A, where the experimental results cannot be calculated efficiently on a classical computer. The output requires summation of weights of perfect matchings of a complex weighted graph, which is a remarkably difficult problem that is \textit{$\#P$-hard} \cite{valiant1979complexity, aaronson2011linear}. The experiment consists of $N$ nonlinear crystals and $M$ optical output paths in total. We call this type of experiments "the crystal network" for the rest of the manuscript. One can experimentally adjust the pump power and phases for every crystal, which allows to change every single weight of the edges of the corresponding graph independently. The crystals are pumped coherently and the pump power is set such that $n$ ($n<N$) crystals can produce photon pairs and higher-order pair creations can be neglected. Then we calculate the $2n$-fold coincidence in $2n$ ($2n<M$) output paths. Now one could ask what is the probability of the $2n$-fold coincidences in one specific $2n$ outputs when all crystals are pumped?

In Fig. \ref{fig:fig4BosonSampling}, we show some examples to answer this question. In the first example, we have in total six output paths ($a-f : M=6$) and nine crystals ($N=9$) from which probabilistically two ($n=2$) produce photon pairs. Now we calculate the 4-fold probability for a subset of four output paths (for example, $a$, $b$, $c$ and $d$ highlighted in orange). With the graph-experimental link, a subset of four outputs in the quantum experiment corresponds to a subset of four vertices in the corresponding graph, depicted in orange shown in Fig. \ref{fig:fig4BosonSampling}A. The experimental outcome corresponds to summing weights of perfect matchings of the sub-graph, which is related to calculating the \textit{Permanent} of sub-matrix of the adjacency matrix\footnote{An adjacency matrix is a square matrix used to represent simple graph. The elements of the matrix stand for the weights of the edges between two vertices.}. Therefore, we find that the probability $P_{abcd}$ is proportional to the \textit{Permanent}, $P_{abcd}\propto |Perm(U_{P_{s}})|^2$.

For experiments with general arrangements of crystals, the $2n$-fold probability can be calculated by a generalization of the \textit{Permanent} -- the so-called \textit{Hafnian} \cite{caianiello1953quantum}, shown in Fig. \ref{fig:fig4BosonSampling}B. When the crystal network consists of a large number of crystals, it is unknown how to efficiently approximate the \textit{Hafnian} \cite{bjorklund2012counting, barvinok2017approximating}. To the best of our knowledge, the fastest algorithm to compute the \textit{Hafnian} of a $n\times n$ complex matrix runs in $O(n^{3}2^{n/2})$ time \cite{bjorklund2018faster}.

The task described above is connected to \textit{BosonSampling} \cite{aaronson2011computational, lund2017quantum, spring2012boson, broome2013photonic, crespi2013integrated, tillmann2013experimental}, which requires the matrix function \textit{Permanent} to calculate the experimental results. However, the experimental implementation is fundamentally different. In \textit{BosonSampling} experiments to date, single photons undergo multiphotonic Hong-Ou-Mandel effect \cite{tillmann2015generalized, menssen2017distinguishability, dittel2018totally} in a passive linear optical network. In contrast to that, our concept is based solely on  probabilistic pair sources where frustrated pair creation occurs. Computing \textit{Hafnians} has only recently been investigated by a complementary approach called Gaussian \textit{BosonSampling} \cite{lund2014boson, hamilton2017gaussian, bradler2017gaussian}.

An interesting question is the scaling of expected count rates of BosonSampling and the approach presented here. In the original BosonSampling proposal, $n$ pairs of heralded single photons from $n$ SPDC crystals (with emission probability $p$) are the input into a linear optical network. The countrates for n-fold coincidences $R$ is $R_{BS} \approx p^n$. Later, two independent groups discovered a method to exponentially increase the count rate, called \textit{Gaussian BosonSampling} and \textit{Scattershot BosonSampling} \cite{lund2014boson, AaronsonBlog}. There, each of the $m$ inputs of the BosonSampling device is feeded with one output of an SPDC crystal (the second SPDC photon is heralded). That means, there are $m$ SPDC crystals ($m > n$). That leads to an exponential increased count rate for n-fold coincidences of $R_{SS} \approx \binom{m}{n} p^n \left(1-p\right)^{m-n}$, which is the input in the BosonSampling device.

Estimating the count rates in our approach needs a slightly more subtle consideration, as our photons are not the input to a BosonSampling device but their generation itself is in a superposition. Let us look at the example given in Fig.~\ref{fig:fig4BosonSampling}\text{A}. Here we compare a complete bipartite graph to scattershot boson sampling. For a complete bipartite graph, we have two sets of paths $\{a,c,e\}$ and $\{b,d,f\}$. To calculate the probability of detecting a four-fold coincidence, we first derive all possible crystal combinations that could lead to a four-fold detection. There are $\binom{3}{2}$ ways to choose two elements from the two sets of paths. Therefore, there exist $\binom{3}{2}\times\binom{3}{2}$ combinations of crystal pairs that produce 4-fold coincidences. In general, for $m^2$ crystals and $2n$-fold coincidendes we have $\binom{m}{n}^2$. Furthermore, each combination can arise due to two (in general $n!$) indistinguishable crystal combinations. For example, a $(abcd)$ four-fold detection can arise either from a photon pair emission from crystals I\&IV or II\&VI, as depicted in Fig.~\ref{fig:fig4BosonSampling}\text{A}. Of course, the relative phase between these possibilities detemine whether we expect constructive or destructive interference. The latter case would not contribute any counts. Since for boson sampling the phases are randomly distributed, we average over a uniform phase distribution to account for all possbile phase settings. This is equivalent to a two-dimensional random walk. Thus in general the average magnitude of the amplitude gives $\sqrt{n!}$. Therefore, the count rate is magnified by $n!$.
%
%
Finally, the estimated count rate for our new approach based on path identity is $R_{PI} \approx \binom{m}{n}^2 n! p^n \left(1-p\right)^{m^2-n}$. The ratio of the Path Identity Sampling and Scattershot BosonSampling thus is $\frac{R_{PI}}{R_{SS}}=\binom{m}{n} n! \left(1-p\right)^{m(m-1)}$. This exponential increase is due to the additional number of crystals (while Scattershot BS uses $m$ crystals, we use $m^2$), and the coherent superposition of $n!$ possibilities to receive the output. We compare now this ratio for two recent experimental implementations of Scattershot BosonSampling. In 2015, a group performed Scattershot BosonSampling with $m=13$ and $n=3$ \cite{bentivegna2015experimental}. With $p\approx 0.01$, our approach would lead to roughly 350 times more 2n-fold count-rate. In 2018, a different group performed Scattershot BosonSampling with $m=12$ and up to $n=5$ \cite{zhong201812}. With the same number of modes and photon pairs, we would expect roughly 25000 more 2n-fold count-rate. In SI Appendix \cite{supp}, we explain the scaling based on an example.

For realistic experimental situations, one needs to carefully consider the influence of multi-pair emissions, stimulated emission, loss of photons (including detection efficiencies) and amount of photon-pair distinguishabilities in connection with statements of computation complexity (such as done, for instance, in \cite{rohde2012error, arkhipov2015bosonsampling, rahimi2016sufficient, wang2018toward}). A full investigation of these very interesting questions is out of scope of the current manuscript.

\subsection*{Linear Optics and Graphs}

\begin{figure}[!t]
\centering
\includegraphics[width=0.75\linewidth]{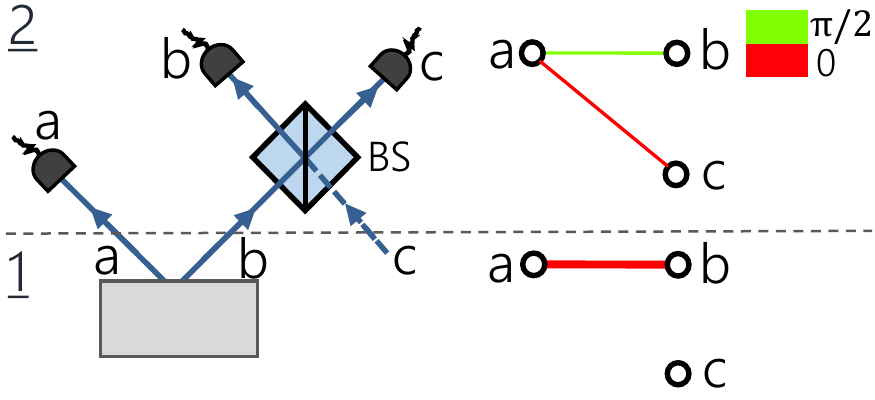}
\caption{The action of a beam splitter described with graph. Here we show a simple linear optical setup with one 50:50 beam splitter. Using graph technique, one can describe the setup as a graph depicted on the right side. Step \textbf{\underline{1}}: A crystal produces a correlated photon pair in paths $a$ and $b$ and no photon goes to path $c$. Therefore there is an edge between vertices $a$ and $b$ and there is no edge connecting vertex $c$. Step \textbf{\underline{2}}: The photon in path $b$ propagates to the beam splitter which will transmit to path $c$ or reflect to path $b$ with an additional phase of $\pi/2$. Therefore, in the case of transmission, the existent red edge $E_{ab}$ will connect the vertex $a$ and $c$. While in the case of reflection, the existent edge $E_{ab}$ gets a complex weight with phase of $\pi/2$ shown in green.}
\label{fig:fig5BeamSplitter}
\end{figure}

\begin{figure}[!t]
\centering
\includegraphics[width=0.83\linewidth]{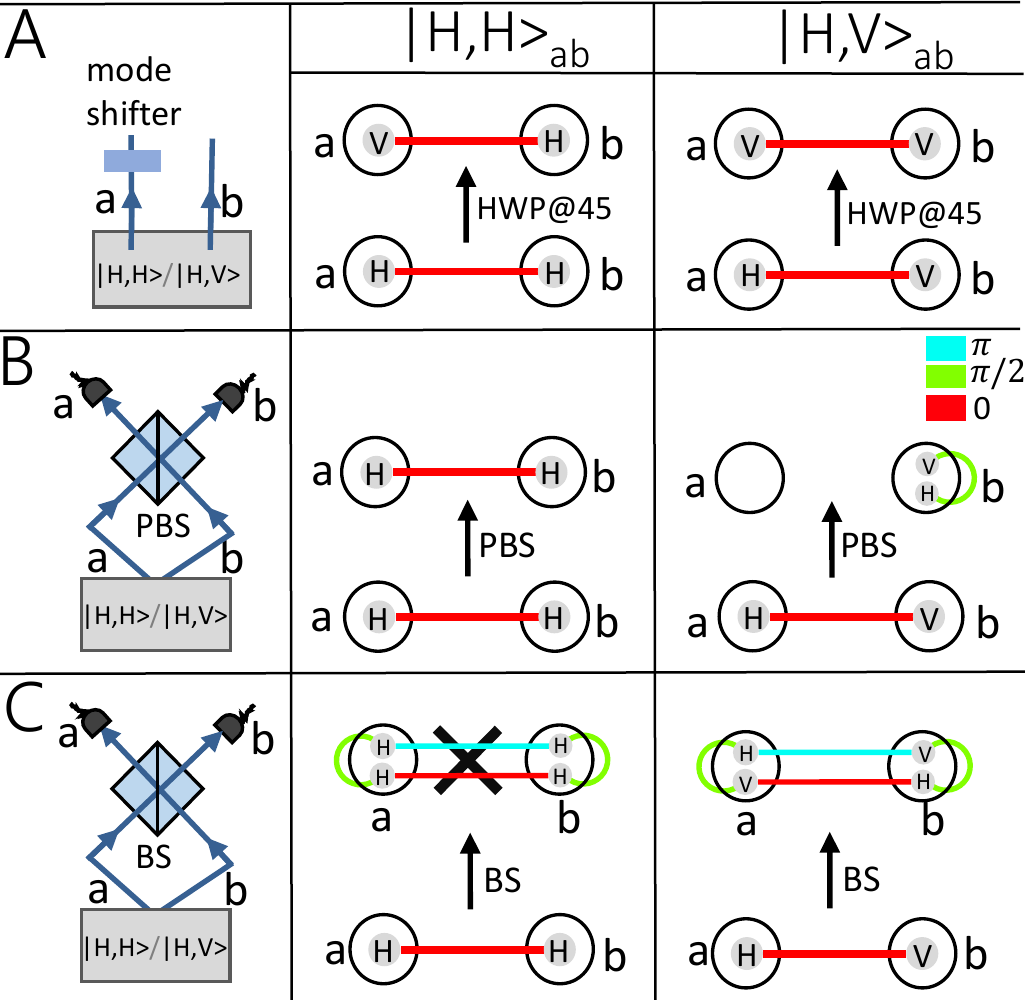}
\caption{\textbf{A}: An example for describing mode shifters with graph. A crystal generates a polarized photon pair in paths $a$ and $b$. A half wave plate (HWP@45) changes the photon's polarization such that horizontal polarization changes to vertical polarization and vice versa. The corresponding graphs are depicted in the right side. The vertices with labels H or V represent horizontal or vertical polarization of photons. Therefore the label H changes to V in the \textit{vertex set a}. \textbf{B}: Graph description for the polarizing beam splitter (PBS). A PBS can transmit horizontally polarized photon and reflect vertically polarized photon. When the crystal creates horizontally polarized photon pair, thus photons in paths $a$ and $b$ transmit to paths $b$ and $a$. When the crystal produces an orthogonally polarized photon pair, photon in path $a$ is transmitted and photon in path $b$ is reflected with phase of $\pi/2$. Therefore there are two correlated photons in path $b$. In the graph, there are two labeled vertex in \textit{vertex sets} $b$ with a green edge connecting them. \textbf{C}: An optical setup for Hong-Ou-Mandel (HOM) interference. A crystal produces a photon pair in paths $a$ and $b$ which propagate to a 50:50 beam splitter. By using the \textit{BS operation}, we get the final graph. Now let's look at two cases where the photons are indistinguishable or distinguishable. For simplicity, we show the example with polarization. When the two photons are indistinguishable -- all of their mode numbers are identical such as $\ket{H,H}_{ab}$, the edges that connect vertices $a$ and $b$ cancel. The remaining green edges with two vertices in \textit{vertex sets} $a$ or two vertices in \textit{vertex sets} $b$ show that there are two photons in path $a$ or $b$. This is a manifestation of the HOM interference. While in the case that the input photons have orthogonal polarization such as $\ket{H,V}_{ab}$, we clearly see that no interference can be observed. Therefore the four possible outputs remain ($\ket{H,V}_{aa}$, $\ket{H,V}_{ab}$, $\ket{V,H}_{ab}$ and $\ket{V,H}_{bb}$).}
\label{fig:fig6HOM}
\end{figure}

\begin{figure*}[!ht]
\centering
\includegraphics[width=\linewidth]{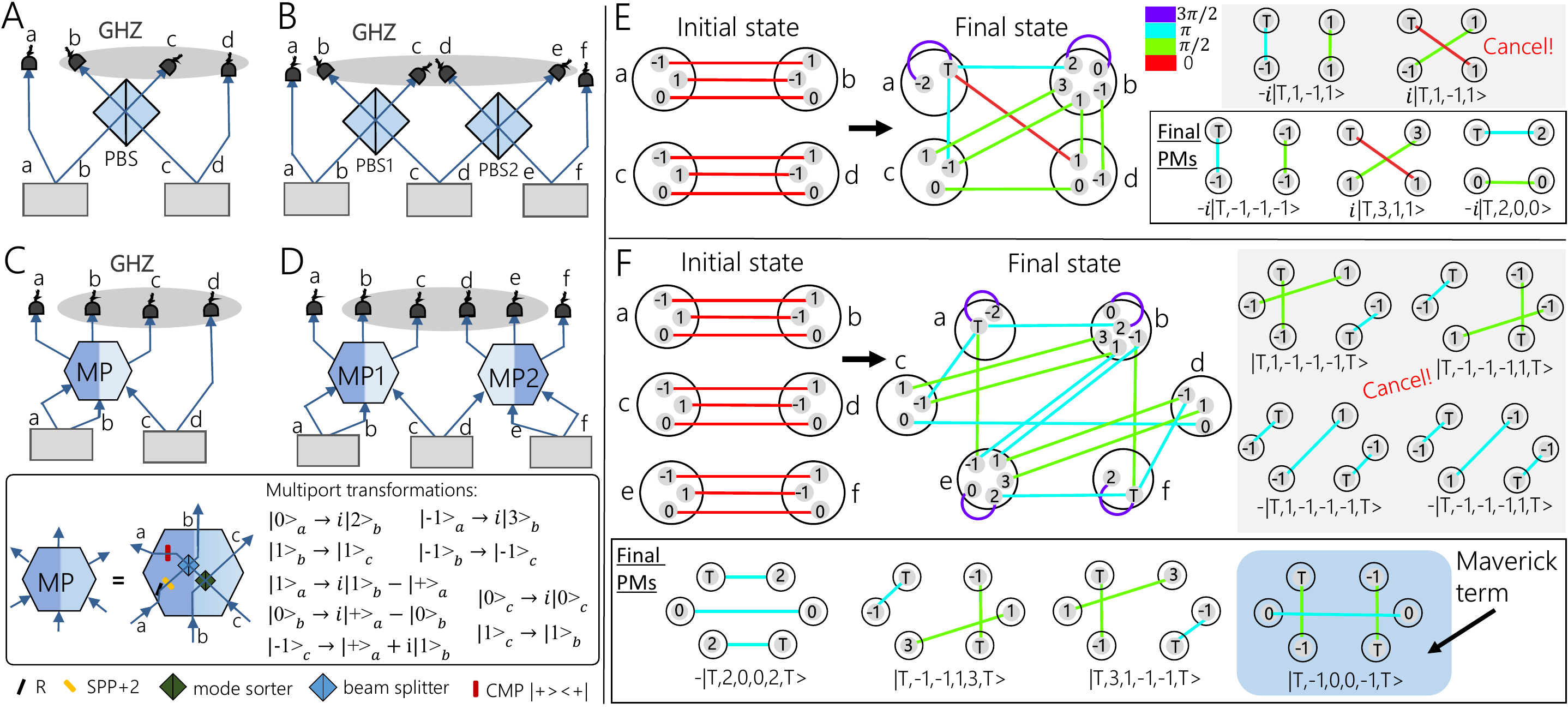}
\caption{Generating high-dimensional multi-photonic states with linear optical setups. \textbf{A}: An optical setup for creating a 2-dimensional 3-photon GHZ state. In this example, each crystal produces an entangled state $\ket{\psi^+}=1/\sqrt{2}(\ket{H,H}+\ket{V,V})$. The photons propagate to a polarizing beam splitter (PBS), and 4-fold coincidences lead to a 2-dimensional 3-photon GHZ state (where the photon in path $a$ acts as a trigger). \textbf{B}: For generating high photon number GHZ states, one can add more crystals and connect them via many PBSs. \textbf{C}: In an analogous way, a 3-dimensional 3-photon GHZ state ($\ket{\psi}=1/\sqrt{3}(\ket{0,0}+\ket{-1,1}+\ket{1,-1})$) has been created recently, by connecting two crystals (each producing a 3-dimensionally entangled photon pair) with a 3-dimensional multi-port (MP) \cite{erhard2018experimental}. The multiport consists of a reflection (R, such as mirrors), a spiral-phase-plate (SPP), a beam splitter, an orbital angular momentum (OAM) mode sorter \cite{leach2002measuring} and a coherent mode-projection (CMP) which projects photon in path a on $\ket{+}=\ket{T}=1/\sqrt{2}(\ket{0}+\ket{-1})$. The corresponding transformation is described under the setup \cite{erhard2018experimental}. \textbf{D}: In order to create higher-dimensional GHZ state, we now want to extend the setup to create a 3-dimensional GHZ state with 4 particles. However, since this setup uses 6 photons, we expect (due to the result in \cite{krenn2017quantum}) to get an additional term in the final quantum state after post-selection. \textbf{E}: The graphs describing the setup in C, where the \textit{vertex set} (large black circle) shows the mode numbers of the photons. The initial state shows three connections for each \textit{vertex set}, which stands for the initial 3-dimensional entanglement (details in the SI Appendix \cite{supp}). The quantum state conditioned on 4-fold coincidences is obtained by calculating the perfect matchings of the graph. There are five perfect matchings and two of them cancel each other, which result in a 3-dimensional GHZ state after triggering the photon in path $a$ on $\ket{T}=1/\sqrt{2}(\ket{0}+\ket{-1})$. \textbf{F}: These graphs describe the experimental setup in D. As expected, it has four perfect matchings (the other four perfect matchings are cancelled), three corresponding to the GHZ state while the fourth one (highlighted in light blue) is the so-called \textit{Maverick term}.}
\label{fig:fig7NogoGHZ}
\end{figure*}
With the complex weights, one can apply the graph method to describe linear optical elements in general linear optical experiments. Firstly, we describe the action of a beam splitter (BS) with our graph language. A crystal produces one photon pair in paths $a$ and $b$ while no photon is in path $c$, as shown in Fig. \ref{fig:fig5BeamSplitter}. Therefore, there is an edge between vertices $a$ and $b$ and there is no edge connecting vertex $c$. The incoming photon from path $b$ propagates to the BS, which gives two possibilities: reflection to path $b$ or transmission to path $c$. In the case of reflection, photons in path $b$ stay in path $b$ with an additional relative phase of $\pi/2$. Thus the correlation between paths $a$ and $b$ will stay and get a relative phase of $\pi/2$. This can be represented as the original red edge keeps connecting vertices $a$ and $b$ while the color of the edge changes to green which stands for a relative phase shift $\pi/2$. In the case of transmission, photons in path $b$ go to path $c$ which changes the original correlation between paths $a$ and $b$ to paths $a$ and $c$. Therefore the original red edge is changed to connect vertices $a$ and $c$.

From the description of the beam splitter above, we can derive the following general rules for BSs, which we called \textit{BS operation} : 1) A BS has two input paths $v$ and $w$,  which corresponds to vertices $v$ and $w$ of the graph. Take one input path $v$ as the start. 2) For transmission, duplicate the existent edges to connect the adjacent vertices of $v$ with vertex $w$ which stands for the other input path of the BS. 3) For reflection, change the colors of the existent edges to the colors which represent a relative phase shift $\pi/2$. 4) Apply step $2$ and $3$ for path $w$.

Another important optical device in photonic quantum experiments is the mode shifter, e.g. half wave plates for polarization or holograms for orbital angular momentum (OAM). The action of mode shifters can also be described within the graph language (see Fig. \ref{fig:fig6HOM}A). The crystal produces an orthogonally or horizontally polarized photon pair in path $a$ and $b$. A mode shifter (such as half wave plates @45) is inserted in path $a$, which will change the photon's horizontal polarization to vertical polarization and vice versa in path $a$. In the graph, we introduce labels for each vertex (small light-gray disks), which indicate the mode numbers of a photon. For example, vertices $a$ and $b$ carry the labels H and V, which stand for the horizontal and vertical polarization. All the mode numbers of one photon in one path are included in a large black circle -- \textit{vertex set}. In the graph language, the operation of a mode shifter can be represented by changing the labels of the vertex.

As another example for the usage of the graph technique, we describe the manipulation of the polarizing beam splitter (PBS) shown in Fig. \ref{fig:fig6HOM}B. In quantum experiments, a PBS transmits horizontally polarized photons and reflects vertically polarized photons with an additional phase of $\pi/2$. If the crystal produces horizontally polarized photon pairs ($\ket{H,H}_{ab}$), photons in path $a$ go to path $b$ and photons in path $b$ go to path $a$. The connection between paths $a$ and $b$ remains. Therefore, the edge between vertices $a$ and $b$ stays as the original red one. If the crystal produces orthogonally polarized photon pairs ($\ket{H,V}_{ab}$), there are two photons in path $b$ -- one photon comes from path $a$ and another photon with an additional phase of $\pi/2$ comes from path $b$ because of reflection. Thus, in the corresponding graph, there are two labeled vertices in \textit{vertex set b} and there is no vertex in \textit{vertex set a}.

Introducing linear optical elements in the graph representation of quantum experiments allows us to describe a prominent quantum effect -- Hong-Ou-Mandel (HOM) interference \cite{hong1987measurement}, which is shown in Fig. \ref{fig:fig6HOM}C. HOM interference can be observed if two indistinguishable photons propagate to different input paths of a beam splitter.

By using the \textit{BS operation}, one can obtain the final graph. When the crystal produces horizontally polarized photon pair, we can immediately see that the edges between \textit{vertex sets} $a$ and $b$ vanish. Thus the experimental setup shows the destructive interference. If the created photons are in orthogonal polarization, the superposition of the perfect matchings is not zero and then no interference can be observed in the experiment.

Every other linear optical elements can be described with graphs. That is because linear optics do not change the number of photons, and cannot destroy photon pair correlations. They can change phases (which changes the complex weight of edges), intrinsic mode numbers (such as polarisation or OAM, which changes the mode number in the vertex set) or the extrinsic mode number (i.e. the path of the photon, which leads to reconnection of edges). All of these actions can be described within our graph method. Thus every linear optical setup with probabilistic photon pair sources corresponds to an undirected graph with complex weights.

Therefore, we are equipped with the powerful technique of the mathematical field of graph theory, which we can now apply to many state-of-the-art photonic experiments.

\subsection*{Restriction for GHZ state generation}
In \cite{krenn2017quantum}, we have shown a restriction on the generation of high-dimensional GHZ states. The limitation stems from the fact that certain graphs with special properties (concerning their perfect matchings) cannot exist. Since we have extended the use of graphs to linear optics, this restriction applies more generally. We show this restriction by investigating a particular linear optical experiment.

To understand this example, let us first analyze the creation of the 2-dimensional GHZ state. For creating a 3-particle GHZ state, we can connect two crystals with a PBS. If the two crystals both create a Bell state, a 3-photonic GHZ state with a trigger in $a$ is created (shown in Fig.\ref{fig:fig7NogoGHZ}A) \cite{pan2001experimental}. Extending this to a 4-particle GHZ state \footnote{A 4-particle polarization GHZ state can also be created in a simpler way by connecting two crystals via a PBS without a trigger with the same setup in Fig.\ref{fig:fig7NogoGHZ}A. However, thereby we emphasis the analogy to the 3-dimensional case.}, we add another crystal that is connected via a PBS as depicted in Fig.\ref{fig:fig7NogoGHZ}B.

Now we are trying exactly the same in a 3-dimensional system. To create a 3-dimensional GHZ state, we can use two crystals (each generating a 3-dimensionally entangled photon pair) and connect them with a 3-dimensional multiport \cite{erhard2018experimental}, as shown in Fig.\ref{fig:fig7NogoGHZ}C. The graphical description for the setup is depicted in Fig.\ref{fig:fig7NogoGHZ}E. There are five perfect matchings of the final graph. When we calculate the sum of the perfect matchings (two of them cancel), we can get the final quantum state written as $\ket{\psi}=\frac{1}{\sqrt{3}}(\ket{3,1,1}-\ket{2,0,0}-\ket{-1,-1,-1})_{bcd}$, which describes a 3-dimensional 3-particle GHZ state\footnote{A 3-dimensional 3-particle GHZ state can be written as $\ket{\psi}=\frac{1}{\sqrt{3}}(\ket{x,y,z}+\ket{\bar{x},\bar{y},\bar{z}}+\ket{\bar{\bar{x}},\bar{\bar{y}},\bar{\bar{z}}})$, where $m\bot \bar{m}\bot \bar{\bar{m}}$ with $m=x,y,z$. The properties of entanglement cannot be changed by local transformations.} \cite{erhard2018experimental}.

In exact analogy to the 2-dimensional case, we add another crystal to the setup, and connect it with another multiport (Fig.\ref{fig:fig7NogoGHZ}D). As in the 2-dimensional case, we would naturally expect to create a 4-particle GHZ state in 3 dimensions with this setup. However, in this setup, 6 photons are used (two triggers and 4 photons for the GHZ state), therefore the corresponding graph has 6 vertices. From \cite{krenn2017quantum} we know that such graphs cannot generate high-dimensional GHZ states because additional terms (so-called \textit{Maverick terms}) occur in the final state\footnote{If the quantum state is independent of the trigger photons, then it consists of only four vertices, and these can be in a 3-dimensional GHZ state. Independent means that edges between the trigger vertices and the state vertices do not appear in any perfect matching.}. And indeed, when we compute the perfect matchings of the graph, the final quantum state with post-selection is given by $\ket{\psi}=\frac{1}{2}(\ket{-1,-1,1,3}-\ket{2,0,0,2}+\ket{3,1,-1,-1}+\ket{-1,0,0,-1})_{bcde}$, which is not a GHZ state because of the additional term $\ket{-1,0,0,-1}_{bcde}$. This is the additional perfect matching that leads to the \textit{Maverick term} (Fig.\ref{fig:fig7NogoGHZ}F), which comes from the tripled photon pairs emission of the middle crystal.

For higher dimensions, even more additional terms will appear -- which can be understood by perfect matchings of graphs. The \textit{Maverick term} is therefore a genuine manifestation of the graph description in a linear optical quantum experiments with a probabilistic photon source. Therefore, 2-dimensional n-particle GHZ states can be created while the 3-dimensional GHZ state with 4 particles is the highest-dimensional entangled GHZ state producable with linear optics and probabilistic photon sources in this way (for instance, without exploiting further ancillary photons).

\begin{figure}[!t]
\centering
\includegraphics[width=0.75\linewidth]{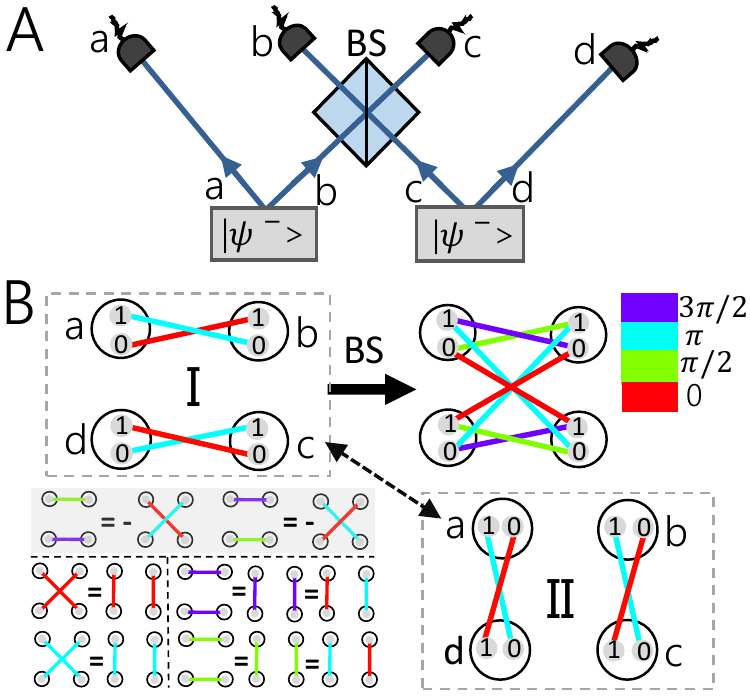}
\caption{Experimental diagram for entanglement swapping and corresponding graph description. \textbf{A}: An experimental setup for entanglement swapping. Each crystal probabilistically generates an entangled state $\ket{\psi^{-}}=\frac{1}{\sqrt{2}}(\ket{0,1}-\ket{1,0})$. When the photons emerge in paths $b$ and $c$ after the beam splitter, the two-photon state in $a$ and $d$ is projected into the Bell state in $\ket{\psi^{-}}$. \textbf{B}: Here we show the experiment using graphs. The initial $\ket{\psi^{-}}$ states (depicted in dotted box \text{I}) both have a relative phase of $\pi$, which is represented by edges with different colors (red and blue). Using the \textit{BS operation}, we get the final graph shown on the right side. There are eight perfect matchings, four of them cancel (highlighted in gray). Due to the symmetry in the quantum state (for example, $\ket{0,0,1,1}_{abcd}$=$\ket{0,1,0,1}_{acbd}$), we rearrange the edges between different vertices after identifying perfect matchings. With $e^{\frac{i3\pi}{2}}e^{\frac{i3\pi}{2}}=e^{i\pi}e^{i2\pi}=e^{i\pi}e^{0}$, perfect matching of two purple edges can be redescribed as one edge in red and another in blue. The perfect matching for green edges is depicted in the similar way. Finally, we obtain the final graph shown in dotted box \text{II}. From the two dotted boxes, we can clearly see the swapping of quantum entanglement.}
\label{fig:fig8EntanglementSwap}
\end{figure}

\subsection*{Graphical description for quantum protocols}
Finally, we show that using graphs can also help for interpreting quantum protocols. In Fig.\ref{fig:fig8EntanglementSwap}, the entanglement swapping is described with graphs \cite{zukowski1993event, pan2012multiphoton}. One crystal produces an entangled state $\ket{\psi^{-}}=\frac{1}{\sqrt{2}}(\ket{0,1}-\ket{1,0})$, which can be rewritten as a superposition of correlation with a phase of $\pi$. Therefore the initial graph has two edges between the \textit{vertex set} $a$ and $b$, and two edges between the $c$ and $d$. With the \textit{BS operation}, we can obtain the final graph. In the end, we obtain all perfect matchings and redraw the graph, which shows the entanglement swapping. The link between graph and quantum experiments offers a graphical way to understand experimental quantum applications such as entanglement swapping.

\section*{Conclusion}
We have presented a connection between linear optical quantum experiments with probabilistic photon pair sources and graph theory. The final quantum state after post-selection emerges as a superposition of graphs (more precisely, as a superposition of perfect matchings). With complex weights in the graphs, we find interference of perfect matchings which describes the interference of quantum states. Equipped with that technique, we identify a novel multiphotonic interference effect and show that calculating the outcome of such an experiment on a classical computer is remarkably difficult. Different from the interference which occurs in the \textit{BosonSampling} experiments with linear optics, the underlying effect in our crystal network is \textit{multiphotonic frustrated photon generation}. It would be exciting to see an actual implementation in a laboratory -- potentially in integrated platforms which allow for \textit{on-chip} photon pair generation \cite{jin2014chip, silverstone2014chip, silverstone2015qubit, krapick2016chip,feng2018chip, wang2018multidimensional, santagati2018witnessing, adcock2018programmable}. While we have shown that the expected n-fold coincidence counts will be larger than in conventional BosonSampling systems, an important question is how these systems compete under realistic experimental situations.

Another important question is how these setups can be applied to tasks in quantum chemistry, such as calculations of vibrational spectra of molecules \cite{huh2015boson, sparrow2018simulating}, or topological indizes of molecules \cite{hosoya2002topological}, or graph theory problems \cite{bradler2018graph}.

So far, we focused on n-fold coincidences with one photon per path, which is directly connected to perfect matchings. A generalised graph description which allows for arbitrary photons per path would also be a very interesting question for future research, which will need to exploit not only perfect matchings, but more general techniques in matching theory.

With this connection, we uncovered novel restrictions on classes of quantum states that can be created using state-of-the-art photonic experiments with probabilistic photon sources, in particular, higher dimensional GHZ states. The graph-experimental link could be used for investigating restrictions of other, much large types of quantum states \cite{huber2013structure, goyeneche2016multipartite}, or could help understanding the (non-) constructability of certain two-dimensional states. Restrictions for the generation of quantum states have been found before, using properties of Fock modes \cite{migdal2014multiphoton} for instance, and it would be interesting whether those two independent techniques could be merged. Also severe restrictions on high-dimensional Bell-state measurements are known \cite{calsamiglia2002generalized}, which limits the application of protocols such as high-dimensional teleportation. The application of the graph-theory-link to such types of quantum measurements would be worthwhile.

As an example, we have shown that entanglement swapping can be understood with graphs. A different graphical representation has been developed to describe quantum processes at a more abstract level \cite{coecke2006kindergarten, abramsky2004categorical}. Furthermore, directed graphs have recently been investigated in order to simplify certain calculations in quantum optics, by representing creation and annihilation operators in a visual way \cite{ataman2014field, ataman2015quantum, ataman2018graphical}. A combination of these pictorial approaches with our methods could hopefully improve the abstraction and intuitive understanding of quantum processes.

In \cite{krenn2017quantum}, we have shown that every experiment (based on crystal configurations as shown in Fig. \ref{fig:fig2PathGraph}) corresponds to an undirected graph and vice versa. It is still open whether for every undirected weighted graph, one can find a linear optical setup without path identification. This is an important question for the design of new experiments.

Our method can conveniently describe linear optical experiments with probabilistic photon sources. It will be useful to understand how the formalism can be extended to other type of probabilistic sources, such as single-photon sources based on weak lasers \cite{zhao2004experimental} or three-photon sources based on cascaded down-conversion \cite{hubel2010direct, hamel2014direct} or in general multiphotonic sources \cite{lahiri2018many}. Can it also be applied to other (non-photonic) quantum systems with probabilistic source of quanta?

A final, very important question is how to escape the restrictions imposed by the graph-theory link. Deterministic quantum sources \cite{santori2002indistinguishable, michler2000quantum, senellart2017high} would need an adaption of the description, and active feed-forward \cite{giacomini2002active, pittman2002demonstration, ma2012quantum} is not known how to be described yet -- can they be described with graphs? What are techniques that cannot be described in the way presented here?

\section*{Acknowledgements}
The authors thank Armin Hochrainer, Johannes Handsteiner and Kahan Dare for useful discussions and valuable comments on the manuscript. X.G. thanks Lijun Chen for support. This work was supported by the Austrian Academy of Sciences (\"OAW), by the Austrian Science Fund (FWF) with SFB F40 (FOQUS). XG acknowledges support from the National Natural Science Foundation of China (No.61771236) and its Major Program (No. 11690030, 11690032), the National Key Research and Development Program of China (2017YFA0303700), and from a Scholarship from the China Scholarship Council (CSC).

\bibliographystyle{unsrt}
\bibliography{refs}

\begin{thebibliography}{}

\bibitem{pan2012multiphoton}
J.W. Pan, Z.B. Chen, C.Y. Lu, H. Weinfurter, A. Zeilinger and M. {\.Z}ukowski,
  Multiphoton entanglement and interferometry. \textit{Reviews of Modern
  Physics} \textbf{84}, 777 (2012).

\bibitem{bouwmeester1999observation}
D. Bouwmeester, J.W. Pan, M. Daniell, H. Weinfurter and A. Zeilinger,
  Observation of three-photon Greenberger-Horne-Zeilinger entanglement.
  \textit{Physical Review Letters} \textbf{82}, 1345 (1999).

\bibitem{yao2012observation}
X.C. Yao, T.X. Wang, P. Xu, H. Lu, G.S. Pan, X.H. Bao, C.Z. Peng, C.Y. Lu, Y.A.
  Chen and J.W. Pan, Observation of eight-photon entanglement. \textit{Nature
  Photonics} \textbf{6}, 225 (2012).

\bibitem{wang2016experimental}
X.L. Wang, L.K. Chen, W. Li, H.L. Huang, C. Liu, C. Chen, Y.H. Luo, Z.E. Su, D.
  Wu, Z.D. Li, H. Lu, Y. Hu, X. Jiang, C.Z. Peng, L. Li, N.L. Liu, Y.A. Chen,
  C.Y. Lu and J.W. Pan, Experimental ten-photon entanglement. \textit{Physical
  Review Letters} \textbf{117}, 210502 (2016).

\bibitem{wang201818}
X.L. Wang, Y.H. Luo, H.L. Huang, M.C. Chen, Z.E. Su, C. Liu, C. Chen, W. Li,
  Y.Q. Fang, X. Jiang, J. Zhang, L. Li, N.L. Liu, C.Y. Lu and J.W. Pan,
  18-qubit entanglement with photon's three degrees of freedom.
  \textit{arXiv:1801.04043} (2018).

\bibitem{eibl2004experimental}
M. Eibl, N. Kiesel, M. Bourennane, C. Kurtsiefer and H. Weinfurter,
  Experimental realization of a three-qubit entangled W state. \textit{Physical
  Review Letters} \textbf{92}, 077901 (2004).

\bibitem{wieczorek2009experimental}
W. Wieczorek, R. Krischek, N. Kiesel, P. Michelberger, G. T{\'o}th and H.
  Weinfurter, Experimental entanglement of a six-photon symmetric Dicke state.
  \textit{Physical Review Letters} \textbf{103}, 020504 (2009).

\bibitem{hiesmayr2016observation}
B. Hiesmayr, M. De~Dood and W. L{\"o}ffler, Observation of four-photon orbital
  angular momentum entanglement. \textit{Physical Review Letters} \textbf{116},
  073601 (2016).

\bibitem{malik2016multi}
M. Malik, M. Erhard, M. Huber, M. Krenn, R. Fickler and A. Zeilinger,
  Multi-photon entanglement in high dimensions. \textit{Nature Photonics}
  \textbf{10}, 248 (2016).

\bibitem{erhard2018experimental}
M. Erhard, M. Malik, M. Krenn and A. Zeilinger, Experimental
  Greenberger--Horne--Zeilinger entanglement beyond qubits. \textit{Nature
  Photonics} 1 (2018).

\bibitem{aaronson2011computational}
S. Aaronson and A. Arkhipov, The computational complexity of linear optics.
  \textit{Proceedings of the forty-third annual ACM symposium on Theory of
  computing} 333 (2011).

\bibitem{rahimi2015can}
S. Rahimi-Keshari, A.P. Lund and T.C. Ralph, What can quantum optics say about
  computational complexity theory?. \textit{Physical Review Letters}
  \textbf{114}, 060501 (2015).

\bibitem{lund2017quantum}
A. Lund, M.J. Bremner and T. Ralph, Quantum sampling problems, BosonSampling
  and quantum supremacy. \textit{npj Quantum Information} \textbf{3}, 15
  (2017).

\bibitem{spring2012boson}
J.B. Spring, B.J. Metcalf, P.C. Humphreys, W.S. Kolthammer, X.M. Jin, M.
  Barbieri, A. Datta, N. Thomas-Peter, N.K. Langford, D. Kundys, J.C. Gates,
  B.J. Smith, P.G. Smith and I.A. Walmsley, Boson sampling on a photonic chip.
  \textit{Science} \textbf{339}, 798 (2013).

\bibitem{broome2013photonic}
M.A. Broome, A. Fedrizzi, S. Rahimi-Keshari, J. Dove, S. Aaronson, T.C. Ralph
  and A.G. White, Photonic boson sampling in a tunable circuit.
  \textit{Science} \textbf{339}, 794 (2013).

\bibitem{tillmann2013experimental}
M. Tillmann, B. Daki{\'c}, R. Heilmann, S. Nolte, A. Szameit and P. Walther,
  Experimental boson sampling. \textit{Nature Photonics} \textbf{7}, 540
  (2013).

\bibitem{crespi2013integrated}
A. Crespi, R. Osellame, R. Ramponi, D.J. Brod, E.F. Galvao, N. Spagnolo, C.
  Vitelli, E. Maiorino, P. Mataloni and F. Sciarrino, Integrated multimode
  interferometers with arbitrary designs for photonic boson sampling.
  \textit{Nature Photonics} \textbf{7}, 545 (2013).

\bibitem{carolan2015universal}
J. Carolan, C. Harrold, C. Sparrow, E. Mart{\'\i}n-L{\'o}pez, N.J. Russell,
  J.W. Silverstone, P.J. Shadbolt, N. Matsuda, M. Oguma, M. Itoh, G.D.
  Marshall, M.G. Thompson, J.C.F. Matthews, T. Hashimoto, J.L. O'Brien and A.
  Laing, Universal linear optics. \textit{Science} \textbf{349}, 711 (2015).

\bibitem{bouwmeester1997experimental}
D. Bouwmeester, J.W. Pan, K. Mattle, M. Eibl, H. Weinfurter and A. Zeilinger,
  Experimental quantum teleportation. \textit{Nature} \textbf{390}, 575 (1997).

\bibitem{wang2015quantum}
X.L. Wang, X.D. Cai, Z.E. Su, M.C. Chen, D. Wu, L. Li, N.L. Liu, C.Y. Lu and
  J.W. Pan, Quantum teleportation of multiple degrees of freedom of a single
  photon. \textit{Nature} \textbf{518}, 516 (2015).

\bibitem{pan1998experimental}
J.W. Pan, D. Bouwmeester, H. Weinfurter and A. Zeilinger, Experimental
  entanglement swapping: entangling photons that never interacted.
  \textit{Physical Review Letters} \textbf{80}, 3891 (1998).

\bibitem{zhang2017simultaneous}
Y. Zhang, M. Agnew, T. Roger, F.S. Roux, T. Konrad, D. Faccio, J. Leach and A.
  Forbes, Simultaneous entanglement swapping of multiple orbital angular
  momentum states of light. \textit{Nature Communications} \textbf{8}, 632
  (2017).

\bibitem{krenn2017quantum}
M. Krenn, X. Gu and A. Zeilinger, Quantum Experiments and Graphs: Multiparty
  States as coherent superpositions of Perfect Matchings. \textit{Physical
  Review Letters} \textbf{119}, 240403 (2017).

\bibitem{krenn2016automated}
M. Krenn, M. Malik, R. Fickler, R. Lapkiewicz and A. Zeilinger, Automated
  search for new quantum experiments. \textit{Physical Review Letters}
  \textbf{116}, 090405 (2016).

\bibitem{krenn2017entanglement}
M. Krenn, A. Hochrainer, M. Lahiri and A. Zeilinger, Entanglement by Path
  Identity. \textit{Physical Review Letters} \textbf{118}, 080401 (2017).

\bibitem{wang1991induced}
L. Wang, X. Zou and L. Mandel, Induced coherence without induced emission.
  \textit{Physical Review A} \textbf{44}, 4614 (1991).

\bibitem{herzog1994frustrated}
T. Herzog, J. Rarity, H. Weinfurter and A. Zeilinger, Frustrated two-photon
  creation via interference. \textit{Physical Review Letters} \textbf{72}, 629
  (1994).

\bibitem{hardy1992source}
L. Hardy, Source of photons with correlated polarisations and correlated
  directions. \textit{Physics Letters A} \textbf{161}, 326--328 (1992).

\bibitem{valiant1979complexity}
L.G. Valiant, The complexity of computing the permanent. \textit{Theoretical
  computer science} \textbf{8}, 189 (1979).

\bibitem{aaronson2011linear}
S. Aaronson, A linear-optical proof that the permanent is\# P-hard.
  \textit{Proc. R. Soc. A} \textbf{467}, 3393--3405 (2011).

\bibitem{raussendorf2001one}
R. Raussendorf and H.J. Briegel, A one-way quantum computer. \textit{Physical
  Review Letters} \textbf{86}, 5188 (2001).

\bibitem{hein2006entanglement}
M. Hein, W. D{\"u}r, J. Eisert, R. Raussendorf, M. Nest and H.J. Briegel,
  Entanglement in graph states and its applications. \textit{quant-ph/0602096}
  (2006).

\bibitem{menicucci2006universal}
N.C. Menicucci, P. Loock, M. Gu, C. Weedbrook, T.C. Ralph and M.A. Nielsen,
  Universal quantum computation with continuous-variable cluster states.
  \textit{Physical review letters} \textbf{97}, 110501 (2006).

\bibitem{menicucci2011graphical}
N.C. Menicucci, S.T. Flammia and P. Loock, Graphical calculus for Gaussian pure
  states. \textit{Physical Review A} \textbf{83}, 042335 (2011).

\bibitem{shchesnovich2018collective}
V. Shchesnovich and M. Bezerra, Collective phases of identical particles
  interfering on linear multiports. \textit{Physical Review A} \textbf{98},
  033805 (2018).

\bibitem{perseguers2009quantum}
S. Perseguers, M. Lewenstein, A. Ac{\'\i}n and J.I. Cirac, Quantum random
  networks. \textit{Nature Physics} \textbf{6}, 539 (2010).

\bibitem{cuquet2009entanglement}
M. Cuquet and J. Calsamiglia, Entanglement percolation in quantum complex
  networks. \textit{Physical Review Letters} \textbf{103}, 240503 (2009).

\bibitem{supp}
Supplementary. .

\bibitem{lahiri2015theory}
M. Lahiri, R. Lapkiewicz, G.B. Lemos and A. Zeilinger, Theory of quantum
  imaging with undetected photons. \textit{Physical Review A} \textbf{92},
  013832 (2015).

\bibitem{caianiello1953quantum}
E.R. Caianiello, On quantum field theory-I: explicit solution of Dyson's
  equation in electrodynamics without use of feynman graphs. \textit{Il Nuovo
  Cimento (1943-1954)} \textbf{10}, 1634 (1953).

\bibitem{bjorklund2012counting}
A. Bj{\"o}rklund, Counting perfect matchings as fast as Ryser.
  \textit{Proceedings of the twenty-third annual ACM-SIAM symposium on Discrete
  Algorithms} 914 (2012).

\bibitem{barvinok2017approximating}
A. Barvinok, Approximating permanents and hafnians. \textit{Discrete Analysis}
  \textbf{2}, (2017).

\bibitem{bjorklund2018faster}
A. Bj{\"o}rklund, B. Gupt and N. Quesada, A faster hafnian formula for complex
  matrices and its benchmarking on the Titan supercomputer. \textit{arXiv
  preprint arXiv:1805.12498} (2018).

\bibitem{tillmann2015generalized}
M. Tillmann, S.H. Tan, S.E. Stoeckl, B.C. Sanders, H. Guise, R. Heilmann, S.
  Nolte, A. Szameit and P. Walther, Generalized multiphoton quantum
  interference. \textit{Physical Review X} \textbf{5}, 041015 (2015).

\bibitem{menssen2017distinguishability}
A.J. Menssen, A.E. Jones, B.J. Metcalf, M.C. Tichy, S. Barz, W.S. Kolthammer
  and I.A. Walmsley, Distinguishability and many-particle interference.
  \textit{Physical Review Letters} \textbf{118}, 153603 (2017).

\bibitem{dittel2018totally}
C. Dittel, G. Dufour, M. Walschaers, G. Weihs, A. Buchleitner and R. Keil,
  Totally Destructive Many-Particle Interference. \textit{Physical Review
  Letters} \textbf{120}, 240404 (2018).

\bibitem{lund2014boson}
A. Lund, A. Laing, S. Rahimi-Keshari, T. Rudolph, J.L. O'Brien and T. Ralph,
  Boson sampling from a Gaussian state. \textit{Physical Review Letters}
  \textbf{113}, 100502 (2014).

\bibitem{hamilton2017gaussian}
C.S. Hamilton, R. Kruse, L. Sansoni, S. Barkhofen, C. Silberhorn and I. Jex,
  Gaussian Boson Sampling. \textit{Physical Review Letters} \textbf{119},
  170501 (2017).

\bibitem{bradler2017gaussian}
K. Br{\'a}dler, P.L. Dallaire-Demers, P. Rebentrost, D. Su and C. Weedbrook,
  Gaussian Boson Sampling for perfect matchings of arbitrary graphs.
  \textit{arXiv:1712.06729} (2017).

\bibitem{AaronsonBlog}
S. Aaronson, Scattershot BosonSampling: A new approach to scalable
  BosonSampling experiments. \textit{Shtetl-Optimized,
  https://www.scottaaronson.com/blog/?p=1579} (2013).

\bibitem{bentivegna2015experimental}
M. Bentivegna, N. Spagnolo, C. Vitelli, F. Flamini, N. Viggianiello, L.
  Latmiral, P. Mataloni, D.J. Brod, E.F. Galv{\~a}o, A. Crespi and  others,
  Experimental scattershot boson sampling. \textit{Science advances}
  \textbf{1}, e1400255 (2015).

\bibitem{zhong201812}
H.S. Zhong, Y. Li, W. Li, L.C. Peng, Z.E. Su, Y. Hu, Y.M. He, X. Ding, W.J.
  Zhang, H. Li and  others, 12-photon entanglement and scalable scattershot
  boson sampling with optimal entangled-photon pairs from parametric
  down-conversion. \textit{arXiv preprint arXiv:1810.04823} (2018).

\bibitem{rohde2012error}
P.P. Rohde and T.C. Ralph, Error tolerance of the boson-sampling model for
  linear optics quantum computing. \textit{Physical Review A} \textbf{85},
  022332 (2012).

\bibitem{arkhipov2015bosonsampling}
A. Arkhipov, BosonSampling is robust against small errors in the network
  matrix. \textit{Physical Review A} \textbf{92}, 062326 (2015).

\bibitem{rahimi2016sufficient}
S. Rahimi-Keshari, T.C. Ralph and C.M. Caves, Sufficient conditions for
  efficient classical simulation of quantum optics. \textit{Physical Review X}
  \textbf{6}, 021039 (2016).

\bibitem{wang2018toward}
H. Wang, W. Li, X. Jiang, Y.M. He, Y.H. Li, X. Ding, M.C. Chen, J. Qin, C.Z.
  Peng, C. Schneider and  others, Toward scalable boson sampling with photon
  loss. \textit{Physical Review Letters} \textbf{120}, 230502 (2018).

\bibitem{leach2002measuring}
J. Leach, M.J. Padgett, S.M. Barnett, S. Franke-Arnold and J. Courtial,
  Measuring the orbital angular momentum of a single photon. \textit{Physical
  Review Letters} \textbf{88}, 257901 (2002).

\bibitem{hong1987measurement}
C.K. Hong, Z.Y. Ou and L. Mandel, Measurement of subpicosecond time intervals
  between two photons by interference. \textit{Physical Review Letters}
  \textbf{59}, 2044 (1987).

\bibitem{pan2001experimental}
J.W. Pan, M. Daniell, S. Gasparoni, G. Weihs and A. Zeilinger, Experimental
  demonstration of four-photon entanglement and high-fidelity teleportation.
  \textit{Physical Review Letters} \textbf{86}, 4435 (2001).

\bibitem{zukowski1993event}
M. Zukowski, A. Zeilinger, M.A. Horne and A.K. Ekert, ''Event-ready-detectors''
  Bell experiment via entanglement swapping. \textit{Physical Review Letters}
  \textbf{71}, 4287 (1993).

\bibitem{jin2014chip}
H. Jin, F. Liu, P. Xu, J. Xia, M. Zhong, Y. Yuan, J. Zhou, Y. Gong, W. Wang and
  S. Zhu, On-chip generation and manipulation of entangled photons based on
  reconfigurable lithium-niobate waveguide circuits. \textit{Physical Review
  Letters} \textbf{113}, 103601 (2014).

\bibitem{silverstone2014chip}
J.W. Silverstone, D. Bonneau, K. Ohira, N. Suzuki, H. Yoshida, N. Iizuka, M.
  Ezaki, C.M. Natarajan, M.G. Tanner, R.H. Hadfield, V. Zwiller, G.D. Marshall,
  J.G. Rarity, J.L. O'Brien and M.G. Thompson, On-chip quantum interference
  between silicon photon-pair sources. \textit{Nature Photonics} \textbf{8},
  104 (2014).

\bibitem{silverstone2015qubit}
J.W. Silverstone, R. Santagati, D. Bonneau, M.J. Strain, M. Sorel, J.L. O'Brien
  and M.G. Thompson, Qubit entanglement between ring-resonator photon-pair
  sources on a silicon chip. \textit{Nature Communications} \textbf{6}, 7948
  (2015).

\bibitem{krapick2016chip}
S. Krapick, B. Brecht, H. Herrmann, V. Quiring and C. Silberhorn, On-chip
  generation of photon-triplet states. \textit{Optics Express} \textbf{24},
  2836 (2016).

\bibitem{feng2018chip}
L.T. Feng, M. Zhang, Y. Chen, G.P. Guo, G.C. Guo, D.X. Daiy and X.F. Ren,
  On-chip transverse-mode entangled photon source. \textit{arXiv:1802.09847}
  (2018).

\bibitem{wang2018multidimensional}
J. Wang, S. Paesani, Y. Ding, R. Santagati, P. Skrzypczyk, A. Salavrakos, J.
  Tura, R. Augusiak, L. Man{\v{c}}inska, D. Bacco, D. Bacco, D. Bonneau, J.W.
  Silverstone, Q. Gong, A. Acin, K. Rottwitt, L.K. Oxenlowe, J.L. O'Brien, A.
  Laing and M.G. Thompson, Multidimensional quantum entanglement with
  large-scale integrated optics. \textit{Science} \textbf{360}, eaar7053
  (2018).

\bibitem{santagati2018witnessing}
R. Santagati, J. Wang, A.A. Gentile, S. Paesani, N. Wiebe, J.R. McClean, S.
  Morley-Short, P.J. Shadbolt, D. Bonneau, J.W. Silverstone and  others,
  Witnessing eigenstates for quantum simulation of Hamiltonian spectra.
  \textit{Science advances} \textbf{4}, eaap9646 (2018).

\bibitem{adcock2018programmable}
J.C. Adcock, C. Vigliar, R. Santagati, J.W. Silverstone and M.G. Thompson,
  Programmable four-photon graph states on a silicon chip.
  \textit{arXiv:1811.03023} (2018).

\bibitem{huh2015boson}
J. Huh, G.G. Guerreschi, B. Peropadre, J.R. McClean and A. Aspuru-Guzik, Boson
  sampling for molecular vibronic spectra. \textit{Nature Photonics}
  \textbf{9}, 615 (2015).

\bibitem{sparrow2018simulating}
C. Sparrow, E. Mart{\'\i}n-L{\'o}pez, N. Maraviglia, A. Neville, C. Harrold, J.
  Carolan, Y.N. Joglekar, T. Hashimoto, N. Matsuda, J.L. O'Brien and  others,
  Simulating the vibrational quantum dynamics of molecules using photonics.
  \textit{Nature} \textbf{557}, 660 (2018).

\bibitem{hosoya2002topological}
H. Hosoya, The topological index Z before and after 1971. \textit{Internet
  Electron. J. Mol. Des} \textbf{1}, 428--442 (2002).

\bibitem{bradler2018graph}
K. Bradler, S. Friedland, J. Izaac, N. Killoran and D. Su, Graph isomorphism
  and Gaussian boson sampling. \textit{arXiv preprint arXiv:1810.10644} (2018).

\bibitem{huber2013structure}
M. Huber and J.I. Vicente, Structure of multidimensional entanglement in
  multipartite systems. \textit{Physical Review Letters} \textbf{110}, 030501
  (2013).

\bibitem{goyeneche2016multipartite}
D. Goyeneche, J. Bielawski and K. {\.Z}yczkowski, Multipartite entanglement in
  heterogeneous systems. \textit{Physical Review A} \textbf{94}, 012346 (2016).

\bibitem{migdal2014multiphoton}
P. Migda{\l}, J. Rodr{\'\i}guez-Laguna, M. Oszmaniec and M. Lewenstein,
  Multiphoton states related via linear optics. \textit{Physical Review A}
  \textbf{89}, 062329 (2014).

\bibitem{calsamiglia2002generalized}
J. Calsamiglia, Generalized measurements by linear elements. \textit{Physical
  Review A} \textbf{65}, 030301 (2002).

\bibitem{coecke2006kindergarten}
B. Coecke, Kindergarten quantum mechanics: Lecture notes. \textit{AIP
  Conference Proceedings} \textbf{810}, 81 (2006).

\bibitem{abramsky2004categorical}
S. Abramsky and B. Coecke, A categorical semantics of quantum protocols.
  \textit{Logic in computer science, 2004. Proceedings of the 19th Annual IEEE
  Symposium on} 415 (2004).

\bibitem{ataman2014field}
S. Ataman, Field operator transformations in quantum optics using a novel
  graphical method with applications to beam splitters and interferometers.
  \textit{The European Physical Journal D} \textbf{68}, 288 (2014).

\bibitem{ataman2015quantum}
S. Ataman, The quantum optical description of three experiments involving
  non-linear optics using a graphical method. \textit{The European Physical
  Journal D} \textbf{69}, 44 (2015).

\bibitem{ataman2018graphical}
S. Ataman, A graphical method in quantum optics. \textit{Journal of Physics
  Communications} \textbf{2}, 035032 (2018).

\bibitem{zhao2004experimental}
Z. Zhao, Y.A. Chen, A.N. Zhang, T. Yang, H.J. Briegel and J.W. Pan,
  Experimental demonstration of five-photon entanglement and open-destination
  teleportation. \textit{Nature} \textbf{430}, 54 (2004).

\bibitem{hubel2010direct}
H. H{\"u}bel, D.R. Hamel, A. Fedrizzi, S. Ramelow, K.J. Resch and T. Jennewein,
  Direct generation of photon triplets using cascaded photon-pair sources.
  \textit{Nature} \textbf{466}, 601 (2010).

\bibitem{hamel2014direct}
D.R. Hamel, L.K. Shalm, H. H{\"u}bel, A.J. Miller, F. Marsili, V.B. Verma, R.P.
  Mirin, S.W. Nam, K.J. Resch and T. Jennewein, Direct generation of
  three-photon polarization entanglement. \textit{Nature Photonics} \textbf{8},
  801 (2014).

\bibitem{lahiri2018many}
M. Lahiri, Many-Particle Interferometry and Entanglement by Path Identity.
  \textit{arXiv preprint arXiv:1802.09102} (2018).

\bibitem{santori2002indistinguishable}
C. Santori, D. Fattal, J. Vu{\v{c}}kovi{\'c}, G.S. Solomon and Y. Yamamoto,
  Indistinguishable photons from a single-photon device. \textit{Nature}
  \textbf{419}, 594 (2002).

\bibitem{michler2000quantum}
P. Michler, A. Kiraz, C. Becher, W. Schoenfeld, P. Petroff, L. Zhang, E. Hu and
  A. Imamoglu, A quantum dot single-photon turnstile device. \textit{Science}
  \textbf{290}, 2282 (2000).

\bibitem{senellart2017high}
P. Senellart, G. Solomon and A. White, High-performance semiconductor
  quantum-dot single-photon sources. \textit{Nature Nanotechnology}
  \textbf{12}, 1026 (2017).

\bibitem{giacomini2002active}
S. Giacomini, F. Sciarrino, E. Lombardi and F. De~Martini, Active teleportation
  of a quantum bit. \textit{Physical Review A} \textbf{66}, 030302 (2002).

\bibitem{pittman2002demonstration}
T. Pittman, B. Jacobs and J. Franson, Demonstration of feed-forward control for
  linear optics quantum computation. \textit{Physical Review A} \textbf{66},
  052305 (2002).

\bibitem{ma2012quantum}
X.S. Ma, T. Herbst, T. Scheidl, D. Wang, S. Kropatschek, W. Naylor, B.
  Wittmann, A. Mech, J. Kofler, E. Anisimova, V. Makarov, T. Jennewein, R.
  Ursin and A. Zeilinger, Quantum teleportation over 143 kilometres using
  active feed-forward. \textit{Nature} \textbf{489}, 269 (2012).

\end{thebibliography}

\clearpage
\subsection*{Multiphoton Quantum Interference}
In the crystal networks the interference stems from multiphoton frustrated photon pair creation. Each nonlinear crystal probabilistically creates photon pairs from spontaneous parametric down-conversion. With the theoretical method presented in \cite{wang1991induced,lahiri2015theory}, the down-conversion creation process can be described as
\begin{align}
\hat{U}_{a,b}=\sum_{n=0}^{\infty}\frac{g^{n}}{n!}(\hat{a}^{\dag}\hat{b}^{\dag}-\hat{a}\hat{b})^{n}\label{eq:SPDC}
\end{align}
where $\hat{a}^{\dag}$, $\hat{b}^{\dag}$ and $\hat{a}$, $\hat{b}$ are creation and annihilation operators in paths $a$ and $b$. The $g$ ($g<<1$) is proportional to the down-conversion rate and pump power, which indicates the probability amplitude of creating one photon-pair per pump pulse. Therefore, the quantum state can be expressed as $\ket{\psi}=\hat{U}_{a,b}\ket{vac}$, where $\ket{vac}$ is the vacuum state. In Fig. 3A of the main text, we show the a generalisation of two photon frustrated down-conversion, where the full quantum state can be described as
\begin{align*}
\ket{\psi}=
&(1-2g^2)\ket{0,0,0,0}+g\ket{0,0,H,H}+g^2\ket{0,0,2H,2H}\\
&+g\ket{0,H,0,H}+\sqrt{2}g^2\ket{0,H,H,2H}+g^2\ket{0,2H,0,2H}\\
&+ge^{i\varphi}\ket{H,0,H,0}+\sqrt{2}g^2e^{i\varphi}\ket{H,0,2H,H}+g\ket{H,H,0,0}\\
&+\textcolor{red}{g^2(1+e^{i\varphi})\ket{H,H,H,H}}+\sqrt{2}g^2\ket{H,2H,0,H}\\
&+g^2e^{2i\varphi}\ket{2H,0,2H,0}+\sqrt{2}g^2e^{i\varphi}\ket{2H,H,H,0}\\
&+g^2\ket{2H,2H,0,0}+O(g^3)
\end{align*}

The complete state up to second order SPDC contains exactly one term (depicted in red), which stands for interferences (i.e. its amplitude changes when the phase $\phi$ changes). No other terms, in particular, no other two-photon terms, show that behaviour. Thus, this phenomenon is a genuine multiphotonic interference effect.

\subsection*{Quantum Experiments for Permanents and Hafnians}
In the main text, we present experimental schemes where the output distributions are related to the computation of the matrix function \textit{Permanent} and its generalization \textit{Hafnian}, which are difficult to calculate. All crystals are pumped coherently and the laser power is set in such a way that two photon pairs are produced.

In Fig. \ref{fig:Perm}, with the graph-experimental link, we represent the setup as a graph $G_{P}$, which can also be interpreted as the adjacent matrix $U_{P}$. By adjusting the pump power, we can change $g$ for the amplitudes shown in Table. \ref{table:pumpg}. The parameters for the phase shifters in the experimental setup are obtained from the complex matrix, shown in Table. \ref{table:Perphase}.

The experimental results of the n-fold coincidences are given by the superposition of the perfect matchings of the graph. We take the four-fold coincidences from paths $a$, $b$, $c$ and $d$ as an example (all 15 combinations are described in Fig. \ref{fig:Perm}D). The probability of the four-fold case $P_{abcd}$ is given by the perfect matchings of the sub-graph $G_{P_{s}}$, which is related to calculating the \textit{Permanent} of the sub-matrix $U_{P_{s}}$.

In Fig. \ref{fig:Haf}, we show the general experimental scheme where the results are given by the generalisation of \textit{Permanent}, namely \textit{Hafnian}. In analogue to Fig. \ref{fig:Perm}A, the probability of the four-fold case $P_{abce}$ is given by the perfect matchings of the sub-graph $G_{H_{s}}$, which is related to calculating the \textit{Hafnian} of the sub-matrix $U_{H_{s}}$. All 15 combinations for the four-fold coincidences are described in Fig. \ref{fig:Haf}D. Table. \ref{table:Hafphase} shows the parameters for the phase shifters which comes from the adjacent matrix $U_{H}$ of the graph $G_{H}$.

\begin{table}[!h]
\centering
\caption{the probability amplitude of creating one photon-pair per pump pulse $g$ in Fig. \ref{fig:Perm}A}
\begin{tabular}{c|c}
crystal & adjust pump power for $g$\\ \hline
I & 0.033\\ \hline
II & 0.028\\ \hline
III & 0.025\\ \hline
IV & 0.037\\ \hline
V & 0.014\\ \hline
VI &0.112\\ \hline
VII & 0.033\\ \hline
VIII &0.038\\ \hline
IX & 0.102\\
\end{tabular}
\label{table:pumpg}
\end{table}

\begin{table}[!h]
\centering
\caption{the phases (rads) of phase shifters in Fig. \ref{fig:Perm}A}
\begin{tabular}{c|c|c|c}
a1& -1.2723&d1& 0      \\ \hline
a2& 0      &d2& 1.07525\\ \hline
a3& 4.61318&d3& 0      \\ \hline
b1& 0      &e1& -1.51327\\ \hline
b2& 6.16389&e2& 0      \\ \hline
b3& 0      &e3& 2.35627\\ \hline
c1& 3.34089&f1&0       \\ \hline
c2& 0      &f2& -2.2519\\ \hline
c3& -6.4828&f3& 0      \\
\end{tabular}
\label{table:Perphase}
\end{table}

\begin{table}[!h]
\centering
\caption{the phases (rads) for phase shifters in Fig. \ref{fig:Haf}A}
\begin{tabular}{c|c|c|c}
a1& -1.2723   &d1&3.02198 \\ \hline
a2& 2.58852   &d2& 0      \\ \hline
a3& -0.227233 &d4& -1.10373 \\ \hline
a4& 2.2519    &e2& 0       \\ \hline
b1& 0         &e3& 0       \\ \hline
b2& 4.65062   &e4& 0       \\ \hline
c1& -1.93299  &f2& 0       \\ \hline
c2& 0.419721  &f3& 0       \\ \hline
c3& 0         &f4& 0       \\
\end{tabular}
\label{table:Hafphase}
\end{table}

\begin{figure*}[!t]
\begin{align*}
U_{P}=&\begin{pmatrix}
    0&0&0& 0.0097 - 0.0315i& 0.0277 - 0.0055i &0.0114 + 0.0218i\\
    0&0&0& -0.1110 + 0.0133i & -0.0367 - 0.0074i &0.0066 + 0.0125i \\
    0&0&0& -0.0024- 0.0382i&-0.0347 + 0.0959i &0.0019 - 0.0328i\\
    0.0097 - 0.0315i& -0.1110 + 0.0133i &-0.0024- 0.0382i& 0&0&0\\
    0.0277 - 0.0055i& -0.0367 - 0.0074i &-0.0347 + 0.0959i& 0&0&0\\
    0.0114 + 0.0218i& 0.0066 + 0.0125i &0.0019 - 0.0328i& 0&0&0
\end{pmatrix}
\end{align*}
\end{figure*}

\begin{figure*}[!h]
\centering
\includegraphics[width=0.88\linewidth]{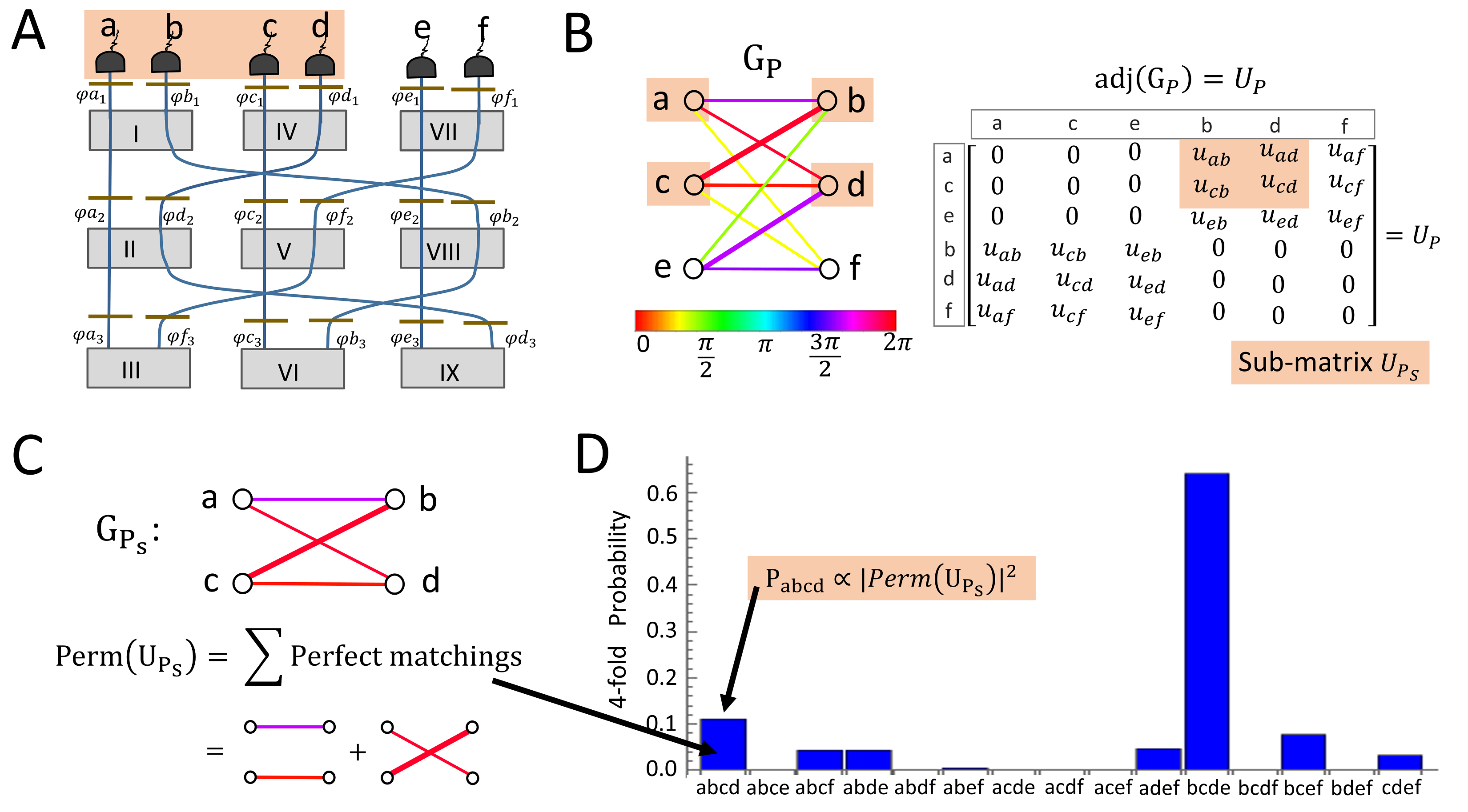}
\caption{Quantum Experiments and the Permanents. \textbf{A}: An experiment consisting of 9 nonlinear crystals (with labels I-IX) and 18 phase shifters (gold lines). They are arranged such that paths $a$, $c$ and $e$ are parallel. All the crystals are pumped coherently and can produce indistinguishable photon pairs. The pump power is set in such a way that two photon pairs are created. One can adjust the phase shifters and pump power to change the phases and transition amplitudes. \textbf{B}: The corresponding graph $G_{P}$ and its adjacency matrix $U_{P}$ for the setup. The ordering of the column and row are ($a$, $c$, $e$, $b$, $d$ and $f$). \textbf{C}: Calculating four-fold coincidences in one specific subset path ($a$, $b$, $c$ and $d$) of four outputs relates to summing weights of the perfect matchings of the sub-graph $G_{P_{s}}$, which relates to the \textit{Permanent} of the sub-matrix $U_{P_{s}}$. Thus, the probability of four-fold coincidences in paths a, b ,c and d $P_{abcd}$ is proportional to the $|Perm(U_{P_{s}})|^2$. \textbf{D}: All the combinations for the four-fold coincidences are depicted in the histogram.}
\label{fig:Perm}
\end{figure*}

\begin{figure*}[h]
\centering
\includegraphics[width=0.9\linewidth]{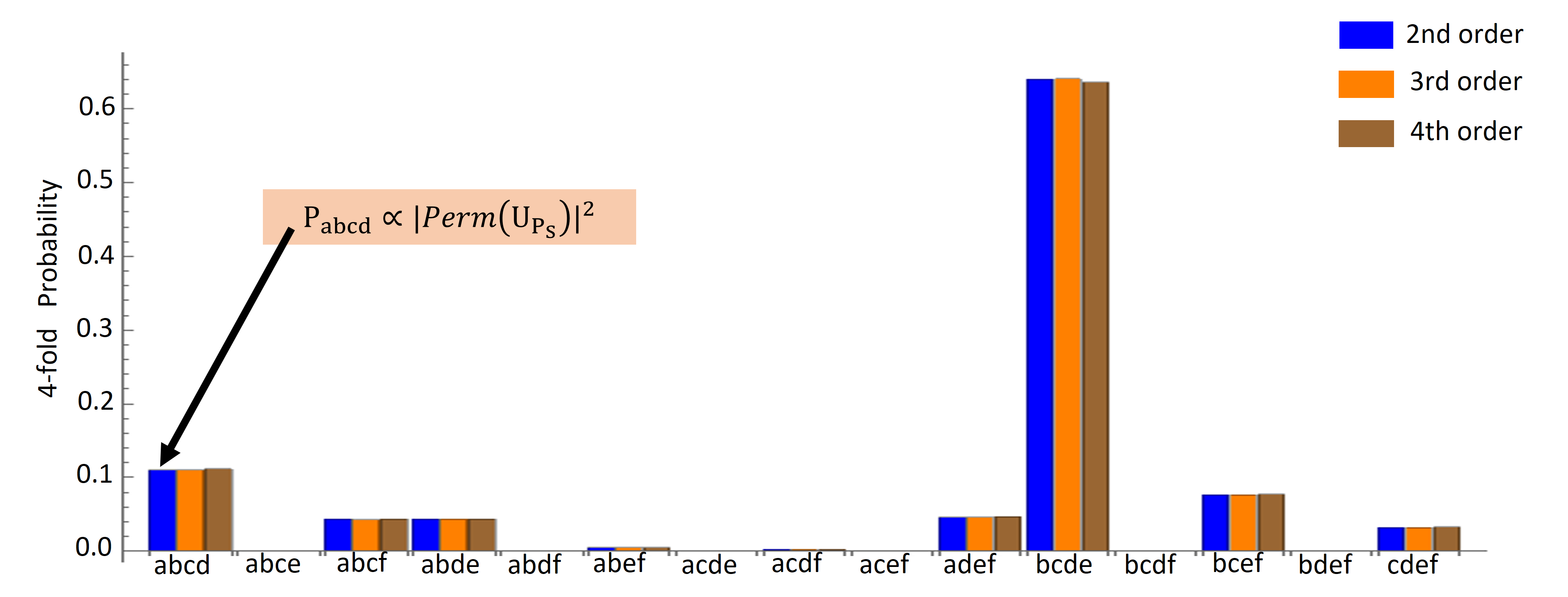}
\caption{Theoretical probabilities for all 15 different four-fold coincidences with 3 and 4-order pair emission, comparing to the 2-order emission in Fig. \ref{fig:Perm}D.}
\label{fig:PefHigh}
\end{figure*}

\begin{figure*}[!t]
\begin{align*}
U_{H}=&\begin{pmatrix}
    0&0.0277-0.0055i&0.0114+0.0218i&0.0097-0.0315i&0 &-0.1110+0.0133i\\
    0.0277-0.0055i&0&0& 0&-0.0367-0.0074i&-0.0024-0.0382i\\
    0.0114+0.0218i&0&0&-0.0347+0.0959&0.0019-0.0328i &0\\
    0.0097-0.0315i&0&-0.0347+0.0959i& 0&0&0\\
    0& -0.0367-0.0074i&0.0019-0.0328i& 0&0&0.0066+0.0125i\\
    -0.1110+0.0133i&-0.0024-0.0382i&0& 0&0.0066+0.0125i&0
\end{pmatrix}
\end{align*}
\end{figure*}

\begin{figure*}[!ht]
\centering
\includegraphics[width=0.88\linewidth]{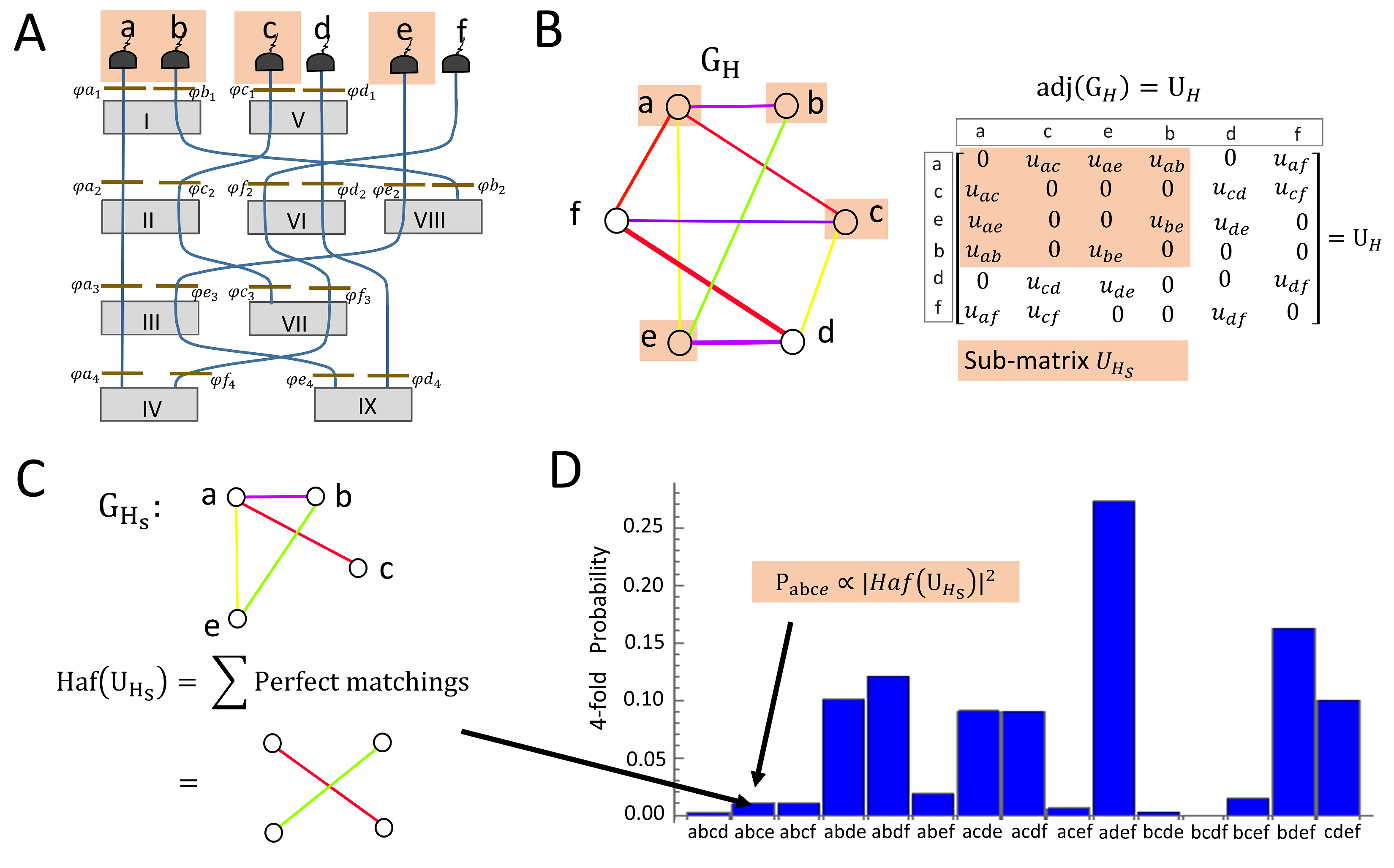}
\caption{Quantum Experiments and the Hafnians. \textbf{A}: A crystal network that shows the general case. The 9 crystals and 18 phase shifters are randomly arranged. In analogous to Fig. \ref{fig:Perm}A, the pump power is also set such that two photon pairs can be produced. \textbf{B}: The corresponding graph $G_{H}$ and its adjacency matrix $U_{H}$ for the setup. The ordering of the column and row are ($a$, $c$, $e$, $b$, $d$ and $f$). \textbf{C}: Again, we take the four-fold coincidence in specific outputs ($a$, $b$, $c$ and $e$) as an example. The result is related to the perfect matchings of the sub-graph $G_{H_{s}}$, which corresponds to computing the \textit{Hafnian} of sub-matrix $U_{H_{s}}$. The probability $P_{abce}$ is given by the matrix function \textit{Hafnian}, $P_{abce} \propto |Haf(U_{H_{s}})|^2$. \textbf{D}: All the 15 combinations for the four-fold coincidence are depicted in the histogram.}
\label{fig:Haf}
\end{figure*}

\begin{figure*}[!t]
\centering
\includegraphics[width=0.88\linewidth]{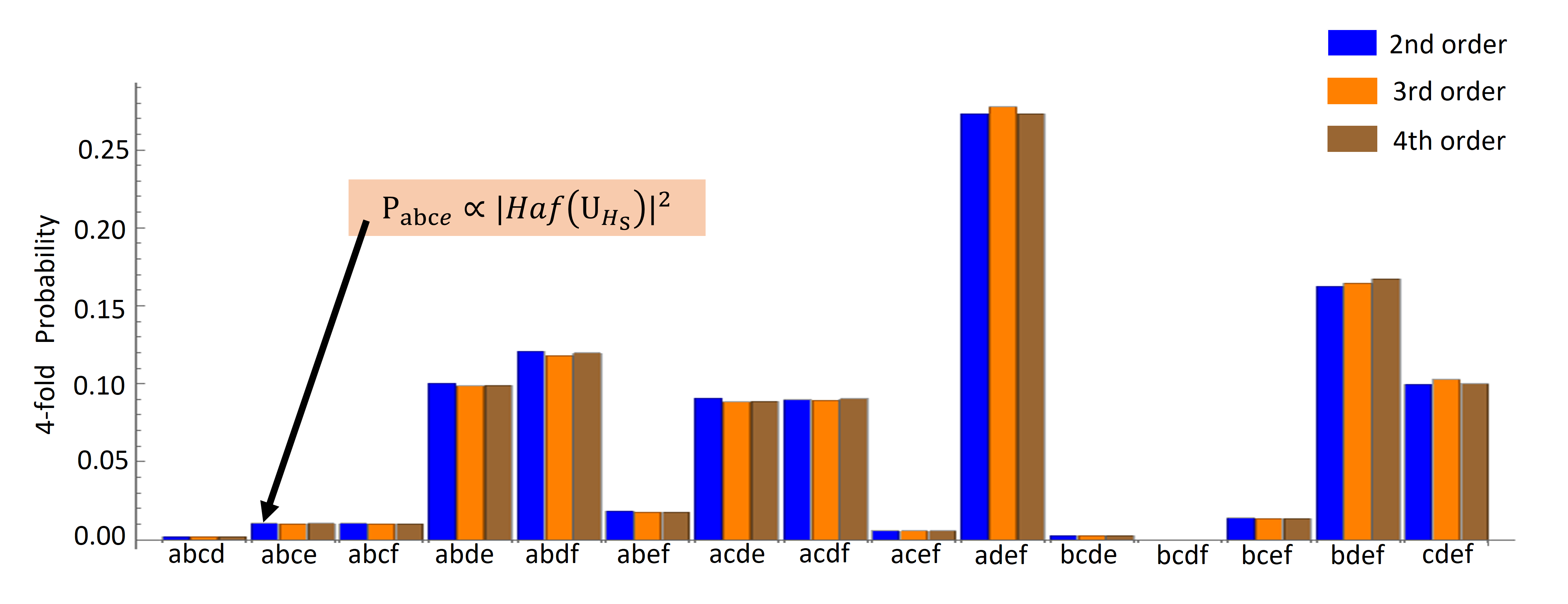}
\caption{Theoretical probabilities for all 15 different four-fold coincidences with 3 and 4-order pair emission, comparing to the 2-order emission in Fig. \ref{fig:Haf}D.}
\label{fig:Hafhigh}
\end{figure*}

\clearpage
\onecolumngrid
\subsection*{Effects of High-order Emission from SPDC and Induced Emission}
Experiments involving probabilistic sources, such as SPDC, exhibit intrinsic error due to higher-order creation processes (see equation \ref{eq:SPDC}). Since $g>0$, there is a possibility that two or more photons in one path. These higher-order terms increase with the laser power, which also contribute to the n-fold coincidences. Here we analyse the influence of this intrinsic error and the expected count rates in the proposed special-purpose quantum computation. Specifically, we analyse the setups and show the influences in Fig. \ref{fig:PefHigh} and Fig. \ref{fig:Hafhigh}.

\begin{figure}[h]
\centering
\includegraphics[width=\textwidth]{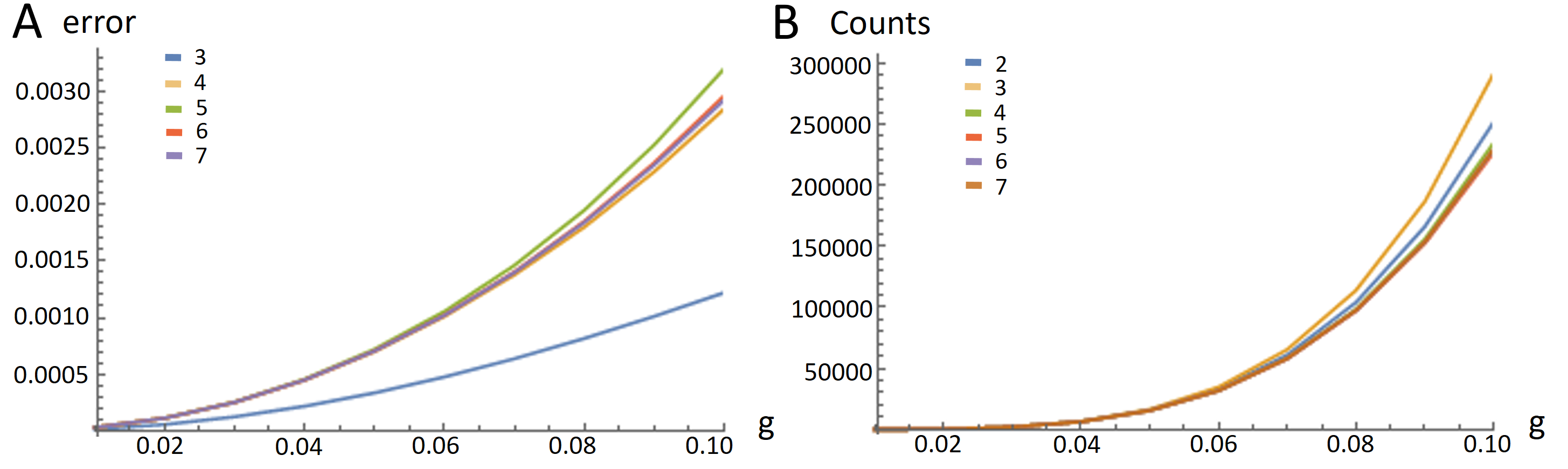}
\caption{Theoretical calculation of the error from induced emission and higher order photon pair emission for the experimental scheme in Fig. \ref{fig:Perm}A. \textbf{A}: There are 9 crystals in the experimental scheme, thus one can adjust the laser power to change the amplitude probability $g$. There are 15 four-fold coincidences cases. For each case, we calculate the scheme with several higher order photon pair emission (3-7) and induced emission. Then the error is given by the average of all the errors for individual cases. The error gets small when the pump power is set weak. \textbf{B}: The theoretical calculation for the count rates of all the 15 combinations with perfect detecting efficiency.}
\label{fig:PerErrorC}
\end{figure}

\begin{figure}[h]
\centering
\includegraphics[width=\textwidth]{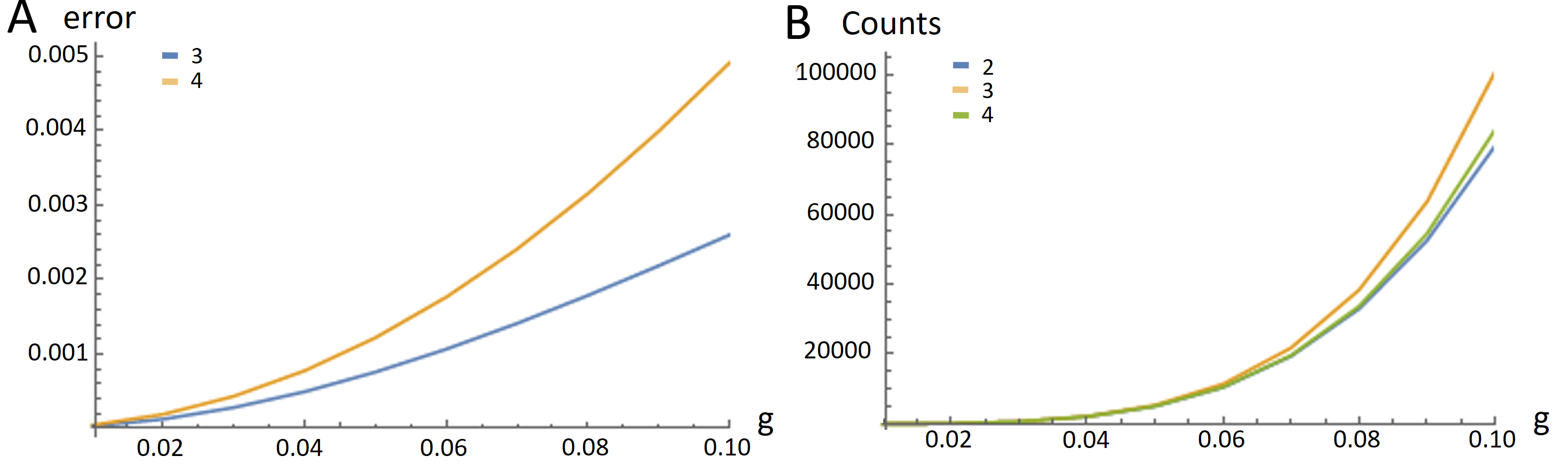}
\caption{In contrast to Fig. \ref{fig:PerErrorC}, we take the photon loss into account. We assume the photon loss is 25$\%$ of the counts. \textbf{A}: The theoretical calculation of 3 and 4-order pair emission is shown. \textbf{D}: The theoretical calculation for the count rates of all the 15 combinations with 25$\%$ loss.}
\label{fig:PerErrorCloss}
\end{figure}

Higher-order photon pair creation is the inherent property of the probabilistic photon source, which can never be removed. However, one can adjust the source power to reduce the influence by making the $g$ to the minimum while keeping enough single-photon count rate. We calculate the error coming from the higher order photon pair generation and induced emission for individual four-fold coincidences case. The error is the average of all the 15 different four-fold coincidences, which is described in Fig. \ref{fig:PerErrorC}A. If one has a pulsed laser with 80MHz repetition rate, then one can get 0.25 million total counts for all the 15 different four-fold coincidence with $g\approx 0.1$, ($p=g^{2}$, which is the probability to produce photon pairs.) see Fig. \ref{fig:PerErrorC}B. However, the detecting and  coupling efficiency are not perfect in the actual experiments. We also theoretically calculate the scheme with photon loss 25$\%$. The error and count rates are described in Fig. \ref{fig:PerErrorCloss}A and B.

\clearpage

\section*{Comparison of count rates for Boson Sampling setups}
\begin{figure}[!ht]
\centering
\includegraphics[width=0.87\textwidth]{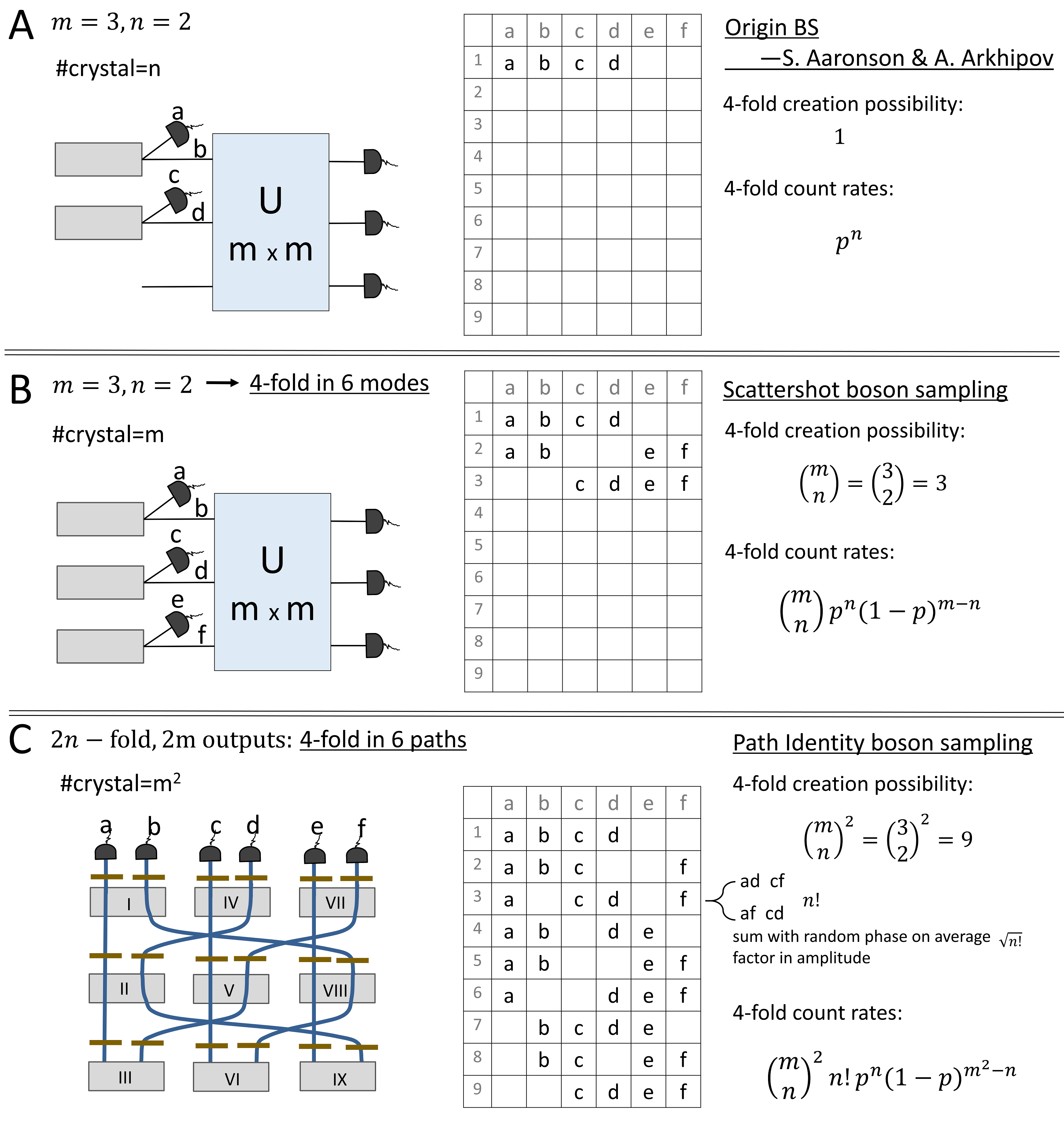}
\caption{Comparison of count rates for three different Boson Sampling schemes. \textbf{A}: Here we show an example to describe the conventional Boson Sampling which was original presented by Scott Aaronson and Alex Arkhipov \cite{aaronson2011computational}. We use $n$ ($n=2$) non-linear crystals to produce photon pairs that $n$ photons are sending into a network of beam splitters and phase shifters, and $n$ photons are heralded. The probability of generating a photon pair from a single source, which is $p\approx g^{2}$. There is only one possibility for 4-fold coincidence. Thus the total count rates of the setup is $p^{n}$. \textbf{B}: Count rates for Scattershot Boson Sampling \cite{lund2014boson}. Different from the Aaronson-Arkhipov Boson Sampling, the number of crystals is the same as the number of the input modes of linear optical network. There are three possibilities for 4-fold coincidence, which is $\binom{m}{n}=\binom{3}{2}=3$. Thus the total count rates are give by $\binom{m}{n}p^{n}(1-p)^{m-n}$.\textbf{C}: Count rates for Path Identity Boson Sampling setup. Our scheme contains $m^{2}$ crystals and there are $\binom{m}{n}^{2}=\binom{3}{2}^{2}=9$ possibilities to produce 4-fold coincidence. There are $n!$ different combinations of the crystals to produce every 4-fold coincidence. The random phases among the crystals distribute randomly. Thus similar to random work, we expect that summing $n!$ random complex numbers leads to an average expected value of $\sqrt{n!}$ (which is the amplitude). Therefore, there is an addition factor of $n!$ in the count rate. The total count rate is $\binom{m}{n}^{2}n!p^{n}(1-p)^{m^{2}-n}$.}
\label{fig:CompareBS}
\end{figure}

In the main text, we present the count rates for three different types of Boson Sampling. Here we explain the count rates with an example described in Fig. \ref{fig:CompareBS}.

\clearpage

\section*{Restrictions for Certain State Generation}
\begin{figure*}[!ht]
\centering
\includegraphics[width=0.95\textwidth]{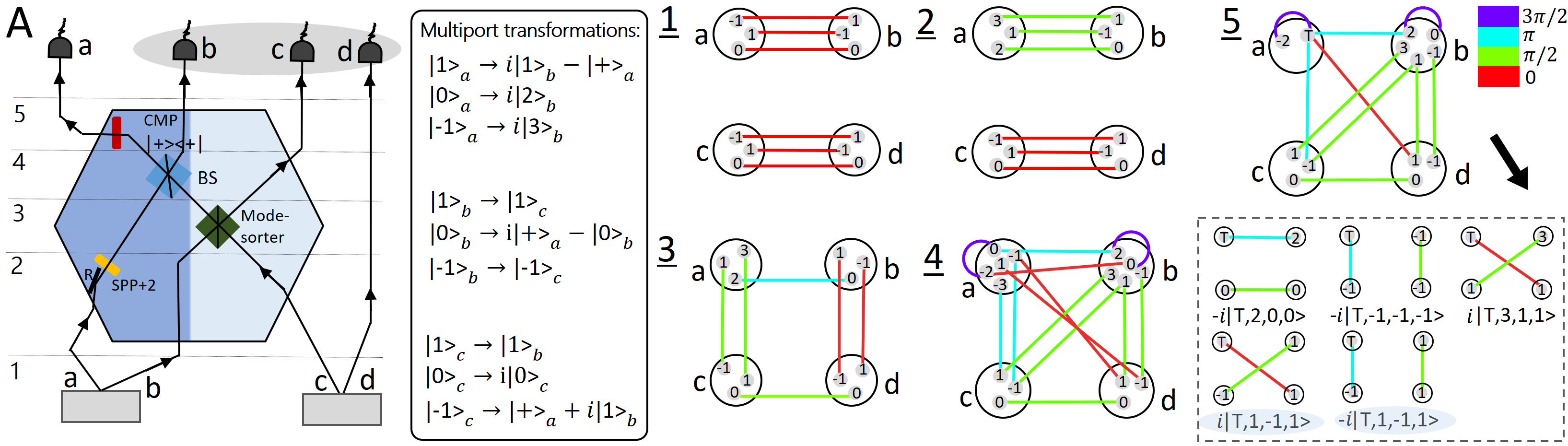}
\caption{Multiport experiment and Graph. \textbf{A}: An experimental setup for producing a 3-dimensional GHZ-state presented in \textit{Experimental Greenberger--Horne--Zeilinger entanglement beyond qubits} \cite{erhard2018experimental}. Each crystal produces a maximally 3-dimensional entangled state $\ket{\psi}=1/\sqrt{3}(\ket{0,0}+\ket{-1,1}+\ket{1,-1})$. The multiport consists of a reflection (R), a spiral-phase-plate (SPP), a beam splitter, an orbital angular momentum (OAM) mode sorter \cite{leach2002measuring} and a coherent mode-projection (CMP) which projects photon in path a on $\ket{+}=\ket{T}=1/\sqrt{2}(\ket{0}+\ket{-1})$. The operation of the multiport is described in the solid box. In the graph, each vertex carries a label which stands for the mode number (such as $0$, $1$, $-1$). The \textit{vertex set} (described with a large gray disk) represents one photon path. Each edge shows a photon pair correlation. The color and width of the edge stands for the phase and probability amplitude. The experiment can be described in the following five steps. \textbf{Step 1}: two crystals produce 3-dimensional 2-photon state between paths $a$ and $b$ and paths $c$ and $d$ respectively. Therefore the initial state is described by three edges connected with \textit{vertex sets} $a$ and $b$ and \textit{vertex sets} $c$ and $d$. \textbf{Step 2}: When the photon propagates through R and SPP, the mode numbers will change as $\ket{\ell}$ $\rightarrow$ $\ket{-\ell+2}$ with an additional phase $\pi/2$. This process can be described by altering the label of vertices in the \textit{vertex set} and the color of related edges. \textbf{Step 3}: The action of OAM sorter in graph. The mode sorter separates incoming photons according to theirs OAM value. Even modes will reflect with a phase $\pi/2$ and even modes will transmit (for example, the mode of a photon in path $c$ propagating to the sorter will change as following: even mode: $\ket{\ell}_c$ $\rightarrow$ $\ket{-\ell}_c$; odd mode: $\ket{\ell}_c$ $\rightarrow$ $\ket{\ell}_b$.). Therefore in the graph, vertices carrying even labels in \textit{vertex set} $c$ will change the sign of the label and the connected edges get an additional complex weight $i$. Vertices carrying the odd labels in \textit{vertex set} $c$ will go to path $b$, therefore their edges $E_{cd}$ are transferred to $E_{bd}$. The photon in path $b$ propagates to OAM sorter in an analogous way. \textbf{Step 4}: When a photon in path $a$ enters the beam splitter, it either reflects to path $a$ with a phase $\pi/2$ or transmits to path $b$. With this \textit{BS operation}, the vertices in $a$ and $b$ will change labels and the original edges would get an additional complex weight $i$. \textbf{Step 5}: The photon in path $a$ will pass through a coherent mode-projection (CMP), which project the state $\ket{0}+\ket{-1}$ into state $\ket{T}$. This is described by changing the labels $0$ or $-1$ of vertex in path $a$ to $T$. Finally, four-fold coincidence counts requires summing weights of perfect matchings of the graph (depicted in dotted box). The quantum state for the experiment is $\ket{\psi}=1/\sqrt{3}(\ket{3,1,1}-\ket{2,0,0}-\ket{-1,-1,-1})_{bcd}$, which is a GHZ state.}
\label{fig:SNogoone}
\end{figure*}

\begin{figure*}[!h]
\includegraphics[width=\textwidth]{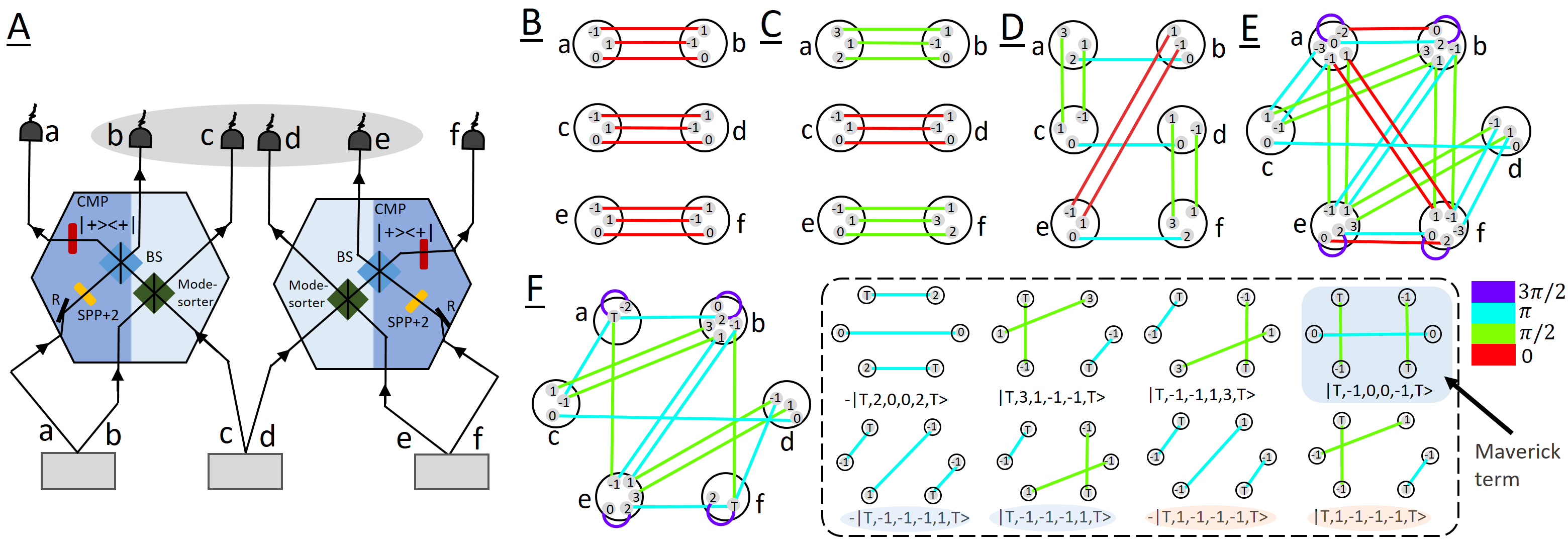}
\caption{Restrictions on creation of quantum states. \textbf{A}: A setup to apparently create a 3-dimensional GHZ-state with 6 particles. Photons in paths $a$, $b$ and $c$ go to MPort1 and others in paths $d$, $e$ and $f$ go to MPort2. This experimental arrangement is analogous to Fig.7B in main text. \textbf{Step 1}: Three crystals produce the 3-dimensional 2-photon pairs in paths $a$, $b$, $c$, $d$, $e$ and $f$. Therefore the initial state is described by six edges connected with the corresponding \textit{vertex sets}. \textbf{Step 2}: The photons in paths $a$ and $f$ go through a $R$ and $SPP$. In the graph language, this operation will change labels of vertices and colors of edges of the graph, respectively. \textbf{Step 3}: Photons in paths $b$ and $c$ propagate to a OAM mode sorter, the same as photons in the paths $d$ and $e$. Similar to Step 3 in Fig.\ref{fig:SNogoone}. \textbf{Step 4}: The photon in path $a$ will reflect to path $a$ with a phase $\pi/2$ or transmit to path $b$, which is similar to photon in path $b$, $e$ and $f$. With this \textit{BS operation}, the labels of the relevant vertices will change and the original edges get an additional complex weight $i$. \textbf{Step 5}: With the projection, we remove all vertices which do not carry the labels ($0$ and $-1$). The triggered vertices are renamed to label $T$. Then, we obtain the final graph of the setup. With six-fold coincidence counts, we calculate the perfect matchings of the graph. There are eight perfect matchings (depicted in dotted box) where four of them cancel. After triggering photons in paths $a$ and $f$, the result state is $\ket{\psi}=1/2(\ket{-1,0,0,-1}-\ket{2,0,0,2}+\ket{3,1,-1,-1}+\ket{-1,-1,1,3})_{bcde}$, which is a 3-dimensional four-photon GHZ-state with the \textit{Maverick term} $\ket{-1,0,0,-1}_{bcde}$.}
\label{fig:SNogotwo}
\end{figure*}

The detailed description of the setup for creating 3-dimensional GHZ state is shown step by step with the graph in Fig.\ref{fig:SNogoone}. Then we show details for the experiment (see Fig7.D in the main text), which is expected to create an 3-dimensional GHZ-state at first sight. However, as known from \cite{krenn2017quantum}, the graph has four perfect matchings, three corresponding to GHZ-state while the fourth one (highlighted in blue) is the so-called \textit{Maverick term}, described in Fig.\ref{fig:SNogotwo}.

\end{document}